\documentclass[oneside,english]{book}
\pdfoutput=1
\usepackage[T1]{fontenc}
\usepackage[latin9]{inputenc}
\usepackage{babel}
\usepackage{bm}
\usepackage{amsmath}
\usepackage{amssymb}
\usepackage{graphicx}
\usepackage{setspace}
\usepackage{esint}
\usepackage[unicode=true,pdfusetitle,
 bookmarks=true,bookmarksnumbered=false,bookmarksopen=false,
 breaklinks=false,pdfborder={0 0 1},backref=false,colorlinks=false]
 {hyperref}

\makeatletter

\numberwithin{equation}{section}
\numberwithin{figure}{section}

\usepackage{braket}

\makeatother

\begin{document}

\begin{titlepage} 
\begin{center}
\textbf{\Huge{}{}Cavity}{\Huge\par}
\par\end{center}

\begin{center}
 
\par\end{center}

\begin{center}
\textbf{\Huge{}{}Optomagnonics }{\Huge\par}
\par\end{center}

\begin{center}
 
\par\end{center}

\begin{doublespace}
\begin{center}
{\huge{}{}Silvia Viola Kusminskiy}{\huge\par}
\par\end{center}
\end{doublespace}

\begin{center}
{\LARGE{}{}Max Planck Institute for the Science of Light}{\LARGE\par}
\par\end{center}

\begin{center}
 
\par\end{center}

\begin{center}
{\large{}{}Staudtstraße 2, 91058 Erlangen, Germany}{\large\par}
\par\end{center}

\begin{center}
 {\large{}{}and}{\large\par}
\par\end{center}

\begin{center}
{\LARGE{}{}Friedrich-Alexander University Erlangen-Nuremberg}{\LARGE\par}
\par\end{center}

\begin{center}
{\large{}{}Staudtstraße 7, 91058 Erlangen, Germany}{\large\par}
\par\end{center}

\begin{center}
 
\par\end{center}

\end{titlepage}

\chapter*{Abstract}

In the recent years a series of experimental and theoretical efforts
have centered around a new topic: the coherent, cavity-enhanced interaction
between optical photons and solid state magnons. The resulting emerging
field of Cavity Optomagnonics is of interest both at a fundamental
level, providing a new platform to study light-matter interaction
in confined structures, as well as for its possible relevance for
hybrid quantum technologies. In this chapter I introduce the basic
concepts of Cavity Optomagnonics and review some theoretical developments.

\tableofcontents{}

\chapter{\label{chap:Introduction}Introduction}

The last two decades have seen enormous advances towards the realization
of quantum technologies \cite{macfarlaneQuantumTechnologySecond2003}.
The ability to bring systems into the quantum regime, to design them,
and to control them, can enable ultra-sensitive measurement and the
manipulation of information at the quantum level, from quantum computers
\cite{aruteQuantumSupremacyUsing2019} to a quantum internet \cite{kimbleQuantumInternet2008}.
At the same time, it permits testing the predictions of quantum mechanics
at unprecedented macroscopic scales \cite{oconnellQuantumGroundState2010a}.

Harnessing the power of quantum mechanics for applications implies
being able to design a system for a certain desired quantum functionality.
In general this means going beyond the single-atom limit, into the
mesoscopic regime. Mesoscopic systems are comprised of millions of
atoms and their behavior is described by collective excitations: e.g.
the mechanical vibrations of a nanobeam. These systems, with characteristic
length scales from tens of nanometers to hundreds of microns, are
such that their collective excitations can be designed and brought
into the quantum regime. This is however very challenging, since quantum
states are fragile and require temperatures lower than the frequencies
of the corresponding collective excitations. Breakthrough experiments
in 2011 used active cooling to bring a macroscopic mode of mechanical
vibration into its quantum ground state \cite{chanLaserCoolingNanomechanical2011a,teufelSidebandCoolingMicromechanical2011a},
by using the backaction of electromagnetic radiation in a cavity.
This is an example of \emph{cavity optomechanical} systems \cite{aspelmeyerCavityOptomechanics2014}.
Very generally, a \emph{cavity} is a ``box'' which serves to confine
the electromagnetic fields and can be used to enhance and even modify
the interaction between electromagnetic radiation and matter.
The field of cavity optomechanics has evolved rapidly, showing, for
example, the possibility of entangling micrometer sized oscillators
via the optomechanical interaction \cite{riedingerRemoteQuantumEntanglement2018}.

Cavity optomechanical systems form part of a broader class of systems
denominated \emph{hybrid quantum systems} \cite{kurizkiQuantumTechnologiesHybrid2015}.
These combine different degrees of freedom, such as photonic, mechanical,
electronic, or magnetic, with the aim of controlling and optimizing
their quantum functionality. For example, while quantum information
can be processed with superconducting qubits at microwave frequencies
\cite{devoretSuperconductingCircuitsQuantum2013}, transmitting the
information through long distances and at room temperature can be
done with optical photons, due to their much higher frequencies. In
turn, storing quantum information requires systems with long coherence
times such that they can act as quantum memories. Promising results
have been obtained in this regard using ensembles of spin impurities
in a solid matrix \cite{afzeliusQuantumMemoryPhotons2015a}. 

In recent years, solid state \emph{magnetic} systems have emerged
as promising candidates for integrating them in hybrid quantum systems.
Current research directions include spintronics \cite{pulizziSpintronics2012},
which aims at using the spin degree of freedom as a carrier replacing
the electron, with the advantage of no energy loss due to Joule heating.
Magnetic and electronic degrees of freedom couple well, and the concept
of current-induced spin torque, proposed theoretically in 1996 \cite{slonczewskiCurrentdrivenExcitationMagnetic1996},
is being nowadays used in the development of random access memories
\cite{bhattiSpintronicsBasedRandom2017}. The field of spin mechanics,
in turn, deals with the coupling of the spin and mechanical degrees
of freedom \cite{losbySpinMechanics2016}. A hybrid spin mechanical
system incorporating also an optomechanical cavity has been moreover
demonstrated for ultra-sensitive magnetometry \cite{wuNanocavityOptomechanicalTorque2017}. 

The interaction between electromagnetic radiation and magnetically
ordered solid state systems in the context of hybrid quantum systems
has however been an unexplored path until quite recently \cite{lachance-quirionHybridQuantumSystems2019a}.
This changed with seminal experiments from 2013 to 2015, in which
\emph{strong coherent coupling} between microwave photons and magnons
in Yttrium Iron Garnet (YIG) was demonstrated \cite{hueblHighCooperativityCoupled2013a,tabuchiHybridizingFerromagneticMagnons2014a,zhangStronglyCoupledMagnons2014a,haighDispersiveReadoutFerromagnetic2015a},
following a theoretical proposal in 2010 \cite{soykalStrongFieldInteractions2010a}.
Magnons are the collective elementary excitations of magnetic systems,
the quanta of the corresponding spin waves in the material. In these 
experiments,
a microwave cavity was used to enhance the spin-photon interaction.
The collective nature of the magnons, involving all spins in the magnetic
sample, also provides a factor of enhancement to the magnon-photon
coupling. The microwave field in the cavity can serve moreover as
an intermediary field to couple the YIG magnons coherently to a superconducting
qubit, indicating the potential of these magnetic systems for quantum
information platforms \cite{tabuchiCoherentCouplingFerromagnetic2015a}.
In 2016, the first experimental \cite{haighTripleResonantBrillouinLight2016,zhangOptomagnonicWhisperingGallery2016,osadaCavityOptomagnonicsSpinOrbit2016}
and theoretical \cite{violakusminskiyCoupledSpinlightDynamics2016,liuOptomagnonicsMagneticSolids2016a}
works on cavity \emph{opto}magnonics appeared, in which an optical
cavity enhances the interaction between the spins and optical photons.

Since MW photons and the probed YIG magnons have similar energies
in the GHz range, the magnon-phonon coupling can be tuned to be resonant,
for example by applying an external magnetic field which controls
the frequency of the magnonic excitations. The coupling term is of
the form
\begin{equation}
g_{{\rm MW}}\left(\hat{S}^{+}\hat{a}^{\dagger}+\hat{S}^{-}\hat{a}\right)\,,\label{eq:MWcoupling}
\end{equation}
written in terms of the spin ladder operators $\hat{S}^{\pm}$ and
the microwave photons operators $\hat{a}^{(\dagger)}$. The first
term creates a photon in mode $\hat{a}$ by annihilating a magnon,
and vice-versa for the second term. The coupling strength between
magnons and photons $g_{{\rm MW}}$ is enhanced with respect to the
single-spin coupling $g_{0}$ due to the collective character of the
magnons by a factor $\sqrt{N}$, where $N$ is the number of spins
participating in the magnon mode \cite{soykalStrongFieldInteractions2010a,lachance-quirionHybridQuantumSystems2019a}.
MW cavity systems with magnetic elements can be by now routinely brought
into the strong coherent coupling regime, with coupling strengths
in the order of hundreds of ${\rm MHz}$ \cite{wangBistabilityCavityMagnon2018,maier-flaigTunableMagnonphotonCoupling2017a,morrisStrongCouplingMagnons2017a,boventerControlCouplingStrength2019,wangQuantumSimulationFermionBoson2019,flowerExperimentalImplementationsCavitymagnon2019}.
In this regime, photons and magnons hybridize, forming a quasiparticle
denominated a \emph{magnon polariton}. Strong coupling is a prerequisite
for quantum information manipulation, since it indicates the rate
at which information is transferred between the different degrees
of freedom. New routes towards tunability and quantum control \cite{lachance-quirionResolvingQuantaCollective2017,raoLevelAttractionLevel2019,boventerSteeringLevelRepulsion2019,wangNonreciprocityUnidirectionalInvisibility2019,maier-flaigTunableMagnonphotonCoupling2017a,lachance-quirionEntanglementbasedSingleshotDetection2019},
and on-chip \cite{morrisStrongCouplingMagnons2017a,houStrongCouplingMicrowave2019,liStrongCouplingMagnons2019}
realizations of these systems are also starting to be explored.

The frequency of optical photons is, on the other hand, in the range
of hundred THz and the coupling to magnons is necessarily parametric,
giving rise to inelastic Brillouin scattering \cite{cottamLightScatteringMagnetic1986}.
In its simplest form, the coupling reads
\begin{equation}
g_{{\rm OP}}\hat{S}_{i}\hat{a}^{\dagger}\hat{a},\label{eq:OMcoupling}
\end{equation}
where $\hat{S}_{i}$ is the $i=x,\,y,\,{\rm or\,}z$ component of
the spin operator and in this case optical photons operators $\hat{a}^{(\dagger)}$.
The spin-photon coupling $g_{{\rm OP}}$ in this regime is inherently
weak. The use of an optical cavity has been predicted to boost the
interaction and, under certain conditions, to allow the system to
enter in the strong coupling regime \cite{violakusminskiyCoupledSpinlightDynamics2016,liuOptomagnonicsMagneticSolids2016a,pantazopoulosPhotomagnonicNanocavitiesStrong2017a,grafCavityOptomagnonicsMagnetic2018a,sharmaOptimalModeMatching2019}.
These conditions are however challenging, and the current experimental
implementations are far from the strong coupling regime. In particular,
given the weakness of the intrinsic interaction, optimal mode matching
between magnon and photonic modes is required. Given the complexity
of structured magnetic systems, this necessitates nontrivial theoretical
design and challenging experimental implementation. The challenge
makes for exciting times in cavity optomagnonics research \cite{pantazopoulosTailoringCouplingLight2018,grafCavityOptomagnonicsMagnetic2018a,almpanisDielectricMagneticMicroparticles2018,osadaOrbitalAngularMomentum2018,osadaBrillouinLightScattering2018,haighSelectionRulesCavityenhanced2018a,pantazopoulosHighefficiencyTripleresonantInelastic2019,hisatomiHelicityChangingBrillouinLight2019a,sharmaOptimalModeMatching2019}.
It is to be expected that the shortcomings in the coupling will be
overcome, opening the door to applications in quantum platforms. For
example, magnons could be used as a quantum transducer, converting
information up or down between MW and telecom photons.

Besides applications, cavity optomagnonics provides a unique setup
in which concepts of cavity quantum electrodynamics (QED) can be applied
to a magnetic system and its excitations. Originating from studying
the electromagnetic radiation emission and absorption properties of
single atoms in a cavity \cite{purcelle.mProceedingsAmericanPhysical1946,harocheCavityQuantumElectrodynamics1989,waltherCavityQuantumElectrodynamics2006},
cavity QED is nowadays a well-established framework to study light-matter
interaction with confined electromagnetic fields. The concepts have
been extended with great success to electromagnetic circuits \cite{schoelkopfWiringQuantumSystems2008}
(circuit QED) and, as mentioned above, optomechanics. The extension
to magnetic systems promises rich physics to discover. 

The following sections cover the basics of cavity optomagnonics, restricted
to the coupling of magnetic systems to photons in the optical domain.
We derive the optomagnonic Hamiltonian starting from the Faraday rotation
of light in magnetized solids and analyze in detail two solvable limits.
In one case, we treat the interaction of light with the homogeneous
magnon mode of the magnetic material (denominated the \emph{Kittel
mode}). For this mode, all spins precess in phase and they can be
treated as a single degree of freedom, consisting of a macrospin.
This allows for treating arbitrary dynamics of the macrospin, including
nonlinear dynamics away for the equilibrium point. In the other case,
we study the coupling to arbitrary magnon modes but restricted to
the spin-wave limit, where only small deviations of the spins from
their equilibrium can be treated. In this limit we can address the
coherent interaction of light with magnetic textures. We will derive
both the quantum Langevin and semiclassical equations of motion for
the coupled open quantum system, and study the spin induced dynamics
due to the light in the cavity. Finally, we will go over a proposal
of a quantum protocol for creating non-classical macroscopic states
of the magnetic system by using light.

\chapter{Optomagnonic Hamiltonian}

\begin{figure}
\begin{centering}
\includegraphics[width=0.8\textwidth]{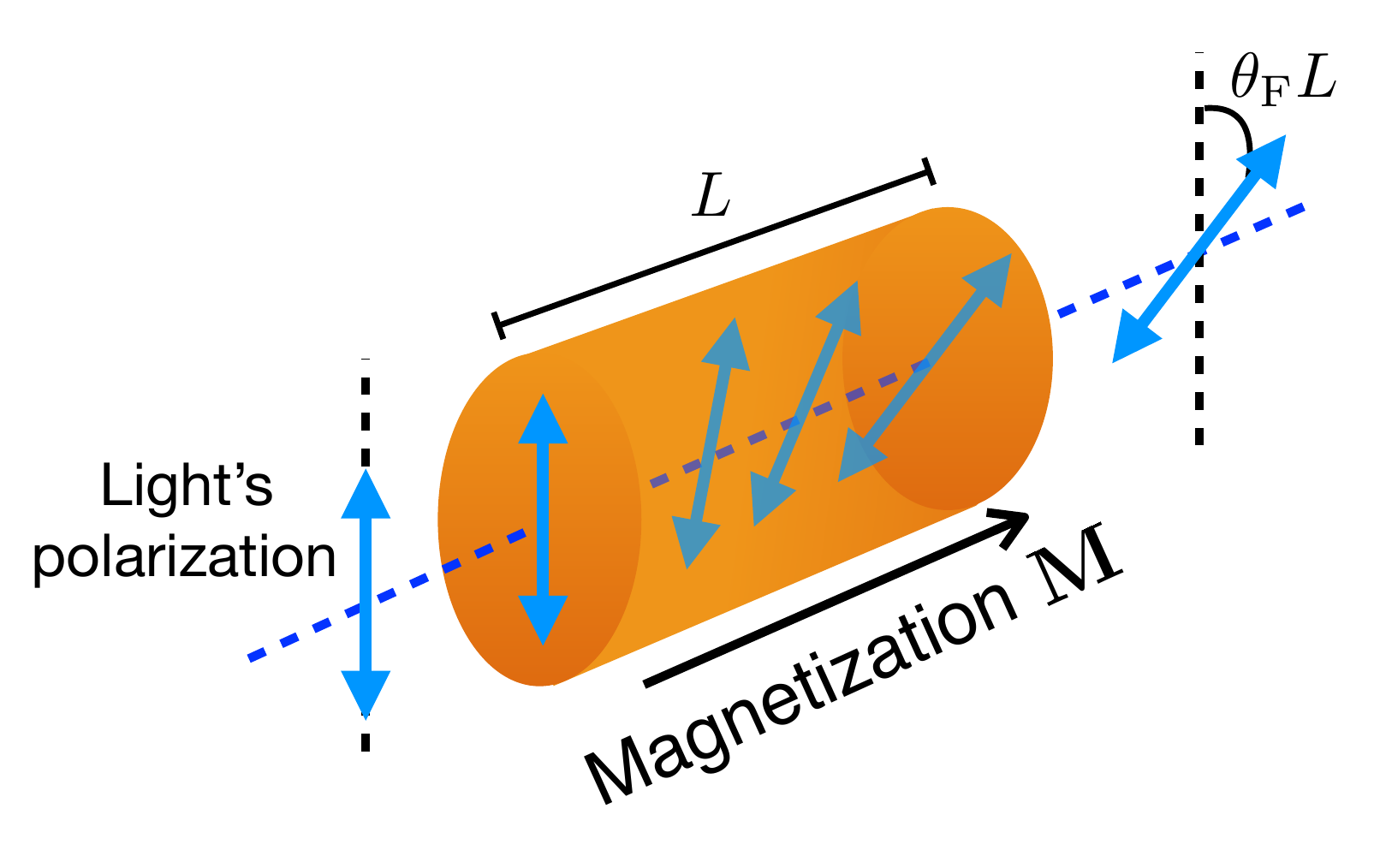}
\par\end{centering}
\caption{Faraday rotation of linearly polarized light propagating through a
magnetized material. The sketch shows the Faraday configuration, where
the light propagates along the magnetization direction. After a length
$L$, the plane of polarization has rotated an angle $\theta_{{\rm F}}L$
in the plane perpendicular to the propagation direction.}
\label{Fig1}
\end{figure}

Cavity optomagnonic systems rely on the interaction between magnetic
insulators and electromagnetic fields at optical frequencies. At these
high frequencies, the magnetic permeability of the material can be
taken as that of the vacuum, $\mu_{0}$, and the magneto-optical interaction
modeled solely through the dielectric permittivity tensor, $\varepsilon_{ij}$
\cite{stancilSpinWavesTheory2009}. One is therefore interested in
the coupling between the electric field component of the electromagnetic
wave, and the magnetization of the host material. This coupling is
responsible for the classical Faraday effect, where the plane of polarization
of light is rotated as the light propagates through the magnetized
medium, see Fig. \eqref{Fig1}. In this configuration, a linearly
polarized plane wave propagates along the magnetization direction,
and the resulting Faraday rotation of the polarization's plane per
unit length of propagation is given by $\theta_{{\rm F}}$, which
is a characteristic of the material (sometimes the Faraday rotation
is given in terms of the \emph{Verdet }constant of the material, which
is the angle of rotation per unit length, per unit magnetic field). 

\section{Faraday Rotation}

The rotation of the plane of polarization of the light can be understood
at a phenomenological level by noting that, if one writes the linear
polarization in a circularly polarized basis, right and left polarizations
inside of the material are not equivalent since time reversal symmetry
is broken due to the magnetization $\mathbf{M}$. Effectively, this
results on a different index of refraction for the two circular polarizations,
which accumulates as a phase difference as light propagates, and results
on the rotation of the polarization in the linear basis. This phenomenon
is also known as \emph{magnetic circular birefringence }and it is
\emph{non-reciprocal}, that is, the acquired phase adds up if the
propagation direction is reversed.

Using the zero energy loss condition, implying that no energy from
the electromagnetic wave is absorbed in the material, together with
the Onsager reciprocity condition for response functions in the presence
of a magnetic field, one can derive symmetry conditions on the permittivity
tensor of the magnetic material. The zero-loss condition is a good
approximation for transparent media. One finds that the permittivity
tensor (i) is Hermitian, (ii) its real part is symmetric in the magnetization,
and (iii) its imaginary part is antisymmetric in the magnetization
\cite{l.d.landauElectrodynamicsContinuousMedia,stancilSpinWavesTheory2009,kusminiskiyQuantumMagnetismSpin2019}.
It is easy to see that a matrix of the form
\begin{equation}
\varepsilon_{ij}(\mathbf{M})=\varepsilon_{0}\left(\varepsilon_{r}\delta_{ij}-if\epsilon_{ijk}M_{k}\right)\label{eq:eps M}
\end{equation}
where $\epsilon_{ijk}$ is the Levi-Civita tensor and $ijk$ are spatial
indices, fulfills conditions (i) to (iii). Throughout this chapter
we use the Einstein convention of summation over repeated indices.
Eq. \eqref{eq:eps M} assumes that the unmagnetized material is isotropic,
this can, however, be readily generalized to non-isotropic materials
by simply replacing $\varepsilon_{r}\delta_{ij}$ by the corresponding
(symmetric and real) permittivity tensor. The linear dependence of
Eq. \eqref{eq:eps M} on the magnetization is valid as long as the correction
of the magnetization on the permittivity is small $\left(fM\ll\varepsilon_{r}\right)$,
which is usually the case. By considering a linearly polarized plane
wave propagating along the material, it is straightforward to show
that the permittivity from Eq. \eqref{eq:eps M} leads to the Faraday
rotation of light (see e.g. Ref. \cite{kusminiskiyQuantumMagnetismSpin2019}).
Denoting the saturation magnetization by $\mathbf{M}_{{\rm s}}$,
$f$ is related to the Faraday rotation coefficient by the expression
\begin{equation}
\theta_{{\rm F}}=\frac{\omega}{2c\sqrt{\varepsilon_{r}}}fM_{{\rm s}}\,\label{eq:theta F}
\end{equation}
where $c=1/\sqrt{\varepsilon_{0}\mu_{0}}$, $\omega$ is the angular
frequency of the light, and the condition $fM_{{\rm s}}\ll\varepsilon_{r}$
has been used.

In optomagnonic systems, one is usually interested in coupling light
to the \emph{excitations} of the magnetically ordered ground state,
and not to the ground state itself. From a classical point of view,
these excitations are time-dependent deviations of the magnetization
with respect to the ground state, in a collective, phase-locked way,
and constitute spin waves. Their respective quanta are denominated
magnons. Hence, in optomagnonics the usual configuration between the
optical fields and the magnetization is not the one from Faraday's
original experiment, where the light propagates along the magnetization
direction, but perpendicular to it. This configuration is denominated
the \emph{Voigt configuration}. The Faraday configuration can however
also be probed and leads to interesting effects concerning angular
momentum conservation rules, as discussed in Ref. \cite{hisatomiHelicityChangingBrillouinLight2019a}
and also in the previous chapter of this book. In \emph{cavity }optomagnonics,
the optical fields form standing waves in a cavity, and part of the
task is to optimize the coupling between the so-called optical spin density
and the magnetic excitations, as we show below.

We are therefore interested in contributions to the permittivity tensor
that are linear in the \emph{deviations} of the magnetization $\mathbf{M}=\mathbf{M}_{0}+\delta\mathbf{M}$,
where $\mathbf{M}_{0}$ is the ground state magnetization and $\delta\mathbf{M}$
the deviation. It is easy to see that terms quadratic in $\mathbf{M}$,
not included in Eq. \eqref{eq:eps M}, also give linear contributions
in $\delta\mathbf{M}$ \cite{cottamLightScatteringMagnetic1986}.
The quadratic contribution in $\mathbf{M}$ to the permittivity gives
rise to the \emph{Cotton-Mouton} effect, also referred to as \emph{magnetic
linear birefringence}. In contrast to the Faraday effect, the Cotton-Mouton
effect is reciprocal. In the simplest situation, the ground state
magnetization is uniform and equal to the saturation magnetization,
$\mathbf{M_{0}=}\mathbf{M}_{{\rm s}}$. Defining the $z$-axis along
$\mathbf{M}_{{\rm s}}$, the deviations of the magnetization can be
written as $\delta\mathbf{M=}(M_{x},M_{y},0)$. In this case, the
only finite components of $\varepsilon_{ij}(\delta\mathbf{M)}$ to
first order in $\delta\mathbf{M}$, including the Cotton-Mouton term,
are given by
\begin{align}
\varepsilon_{yz}(\delta\mathbf{M}) & =\varepsilon_{zy}^{*}(\delta\mathbf{M})=-i\varepsilon_{0}fM_{x}+2\varepsilon_{0}g_{44}M_{y}M_{{\rm s}}\label{eq:eps CM}\\
\varepsilon_{xz}(\delta\mathbf{M}) & =\varepsilon_{zx}^{*}(\delta\mathbf{M})=i\varepsilon_{0}fM_{y}+2\varepsilon_{0}g_{44}M_{x}M_{{\rm s}}\nonumber 
\end{align}
where $g_{44}$ is the Cotton-Mouton coefficient related to the corresponding
rotation angle per unit length 
\begin{equation}
\theta_{{\rm CM}}=\frac{\omega}{2c\sqrt{\varepsilon_{r}}}g_{44}M_{{\rm s}}^{2}\,,\label{eq:theta CM}
\end{equation}
which can be obtained via linear birefringence measurements \cite{stancilSpinWavesTheory2009}.
In the following we will discuss the optomagnonic coupling focusing
only on the Faraday term, considering a permittivity tensor of the
form given by Eq. \eqref{eq:eps M}. The extension to include Cotton-Mouton
terms is however straightforward. From Eq. \eqref{eq:eps CM} one
can see that these terms introduce an asymmetry in the couplings \cite{liuOptomagnonicsMagneticSolids2016a}.

\section{Optomagnonic coupling \label{sec:Magneto-optical-energy}}

If we consider a medium in the presence of time-dependent electromagnetic
fields, we can define an internal energy by taking the time average
of the instantaneous energy. For non-dispersive media the permittivity
tensor is independent of frequency, and it can be shown that the averaged
internal energy can be written as \cite{l.d.landauElectrodynamicsContinuousMedia}
\begin{equation}
E_{{\rm EM}}=\frac{1}{4}\int{\rm d}V\sum_{ij}\left(\mathbf{E}_{i}^{*}\varepsilon_{ij}\mathbf{E}_{j}+\mathbf{H}_{i}^{*}\mu_{ij}\mathbf{H}_{j}\right)\,,\label{eq:Eem}
\end{equation}
where $\mathbf{E}$ is the electric field, $\mathbf{H}$ the magnetic
field, $\mu_{ij}$ the permittivity tensor, and we have used the complex
representation of the fields, i.e. such that ${\rm Re}\left\{ \mathbf{E}\right\} =1/2\left(\mathbf{E}^{*}+\mathbf{E}\right)$.
It is straightforward to show that the expression Eq. \eqref{eq:MO energy}
is real. The magnetization-dependent part of the permittivity introduces
a correction to the electromagnetic energy. Considering for simplicity
only the Faraday rotation term in the permittivity (see Eq. \eqref{eq:eps M}),
from Eq. \eqref{eq:Eem} it is straightforward to show that this correction
is given by

\begin{equation}
U_{{\rm MO}}=-\frac{i}{4}f\varepsilon_{0}\int{\rm d}\mathbf{r}\mathbf{M}(\mathbf{r})\cdot\left[\mathbf{E}^{*}(\mathbf{r})\times\mathbf{E}(\mathbf{r})\right]\,.\label{eq:MO energy}
\end{equation}
The cross product term is proportional to the \emph{optical spin density}
\begin{equation}
\mathbf{S}_{{\rm light}}(\mathbf{r})=\frac{\varepsilon_{0}}{2i\omega}\left[\mathbf{E}^{*}(\mathbf{r})\times\mathbf{E}(\mathbf{r})\right]\label{eq:OSD}
\end{equation}
implying the need of a non-trivial polarization of the optical field
to obtain a finite coupling term. 

One is usually interested in problems where the magnetization $\mathbf{M}$
has a dynamical part, $\mathbf{M}(\mathbf{r},t)=\mathbf{M}_{0}(\mathbf{r})+\delta\mathbf{M}(\mathbf{r},t)$,
where $\delta\mathbf{M}(\mathbf{r},t)$ is the term due to the spin-wave excitations. The static ground state can be uniform ($\mathbf{M}_{0}(\mathbf{r})\equiv\mathbf{M}_{{\rm s}}$)
if the sample is saturated by an external magnetic field, or also
in the case of nanometer samples, where the exchange interaction is
predominant and ---for a ferromagnetic interaction---, sufficient to align
all spins. In general, the interplay of exchange interactions and
dipole-dipole interactions (or other type of interactions, such as
Dzyalozinskii-Moriya \cite{garstCollectiveSpinExcitations2017a})
gives rise either to the formation of domains in macroscopic samples,
or to textured ground states for intermediate sizes (typically microns)
\cite{dennisDefiningLengthScales2002}, where the surface to volume
ratio is large enough as to make the boundaries relevant for the minimization
of the magnetostatic energy \cite{kittelPhysicalTheoryFerromagnetic1949}.
The formation of domains or textures have an exchange energy cost,
but minimize stray fields. The relevant term for the optomagnonic coupling is
the dynamical part of the magnetization $\delta\mathbf{M}(\mathbf{r},t)$.
Therefore Eq. \eqref{eq:MO energy} implies that, besides a nontrivial
polarization of the optical fields, the symmetry of the modes should
be such that the integral is finite. In particular, the overlap between
magnetic and optical modes should be maximized.

Quantizing Eq. \eqref{eq:MO energy} leads to the optomagnonic coupling
Hamiltonian for ferromagnets \cite{violakusminskiyCoupledSpinlightDynamics2016}.
The electric fields $\mathbf{E}(\mathbf{r},t)$ can be readily quantized
in terms of bosonic creation $\hat{a}_{\xi}^{\dagger}$ and annihilation
operators $\hat{a}_{\xi}$,
\begin{equation}
\mathbf{E}(\mathbf{r},t)\rightarrow\hat{\mathbf{E}}(\mathbf{r},t)=\frac{1}{2}\sum_{\text{\ensuremath{\xi}}}\left[\mathbf{E}_{\xi}(\mathbf{r})\hat{a}_{\xi}(t)+\mathbf{E}_{\xi}^{*}(\mathbf{r})\hat{a}_{\xi}^{\dagger}(t)\right]\label{eq:EfieldQuant}
\end{equation}
where $\xi$ labels the corresponding optical mode including the polarization
index. The mode functions $\mathbf{E}_{\xi}(\mathbf{r})$ satisfy
the Helmholtz equation
\begin{equation}
\left(\nabla^{2}+n^{2}k_{0}^{2}\right)\mathbf{E}_{\xi}(\mathbf{r})=0\,,\label{eq:Helmholtz}
\end{equation}
where $n$ is the index of refraction of the medium and $k_{0}$ the
vacuum wave vector. The mode functions are found from Eq. \eqref{eq:Helmholtz}
together with appropriate boundary conditions for the geometry and
material of the optical cavity \cite{heebnerOpticalMicroresonatorsTheory2008}.
The photon operators obey the usual bosonic commutation rules $\left[\hat{a}_{\xi},\hat{a}_{\xi'}^{\dagger}\right]=\delta_{\xi\xi'}$
($\delta_{\xi\xi'}$ the Dirac delta) and $\left[\hat{a}_{\xi},\hat{a}_{\xi'}\right]=\left[\hat{a}_{\xi}^{\dagger},\hat{a}_{\xi'}^{\dagger}\right]=0$. 

The magnetization, in turn, can be quantized in terms of local spin
operators $\hat{\mathbf{s}}_{\mathbf{r}}$ (\textbf{$\mathbf{r}$}
indicates the position of the spin) fulfilling locally the angular
momentum algebra $\left[\hat{\mathbf{s}}_{\mathbf{r}}^{i},\hat{\mathbf{s}}_{\mathbf{r}}^{j}\right]=i\hbar\epsilon_{ijk}\hat{\mathbf{s}}_{\mathbf{r}}^{k}$
and commuting otherwise. The spin operators can be written exactly
in terms of bosonic ones via a Holstein-Primakoff transformation.
In order to preserve the algebra, however, this transformation is necessarily
nonlinear and introduces extra interaction terms between magnons \cite{kusminiskiyQuantumMagnetismSpin2019}.
There are two cases that one can treat, up to a certain extent, analytically:
(i) the uniform case, in which the ground state is uniform and one
is interested in the homogeneous, $\mathbf{k}=0$ spin-wave mode,
denominated the Kittel mode, and (ii) the general, spatial-dependent
case in the spin-wave limit, valid for small deviations of the spins
from their equilibrium configuration. We give the resulting form of
the optomagnonic Hamiltonian for the two cases in the following.

\subsection{Homogeneous magnon mode}

We start with the homogeneous case (i), where both the ground state
magnetization and the excitation are spatially independent: $\mathbf{M}(\mathbf{r},t)=\mathbf{M}_{{\rm s}}+\delta\mathbf{M}(t)$,
with $\mathbf{M}_{{\rm s}}$ the uniform saturation magnetization.
In this case, the magnetization can be quantized simply in terms of
a macrospin $\hat{\mathbf{S}}$
\begin{equation}
\frac{\mathbf{M}}{M_{{\rm s}}}\rightarrow\frac{\hat{\mathbf{S}}}{S}\,,\label{eq:MgotoS}
\end{equation}
where $S$ is the total spin of the considered system. This quantization
scheme allows to retain the spin algebra and to treat fully the nonlinearity
of the problem ($\delta\mathbf{M}(t)$ does not need to be small),
but it is restricted to the homogeneous case. From Eqs. \eqref{eq:EfieldQuant}
and \eqref{eq:MO energy}, using the substitution rule Eq. \eqref{eq:MgotoS},
the optomagnonic coupling Hamiltonian in this case reduces to \cite{violakusminskiyCoupledSpinlightDynamics2016}
\begin{equation}
\hat{H}_{MO}=\hbar\sum_{j\beta\gamma}\hat{S}_{j}G_{\beta\gamma}^{j}\hat{a}_{\beta}^{\dagger}\hat{a}_{\gamma}\label{eq:OM Ham coupling}
\end{equation}
with coupling constants 
\begin{equation}
G_{\beta\gamma}^{j}=-i\frac{\varepsilon_{0}f\,M_{{\rm s}}}{4\hbar S}\xi\epsilon_{jmn}\int{\rm d}\mathbf{r}E_{\beta m}^{*}(\mathbf{r)}E_{\gamma n}(\mathbf{r})\,,\label{eq:G_jbg-1}
\end{equation}
where the Greek indices label the optical modes, and the Roman indices
label the spatial components ($x,\,y,\,z$). The factor $\xi\le1$
is a measure of the overlap between the Kittel mode and the corresponding
optical modes, with $\xi=1$ corresponding to optimal
mode-matching. The $G^{j}$ are hermitian matrices which in general
cannot be simultaneously diagonalized. One sees that there are two
possible kinds of processes: intra-mode coupling, given by the diagonal
elements of $G^{j}$, and inter-mode coupling, given by the off-diagonal
elements. 

The coupling constants $G_{\beta\gamma}^{j}$ are uniquely determined
once the normalization of the electromagnetic field is specified.
We follow the normalization procedure common in optomechanical systems,
where the electromagnetic field amplitude is normalized to one photon
over the EM vacuum \cite{aspelmeyerCavityOptomechanicsNano2014}.
The normalization condition is given by $\hbar\omega_{\alpha}=\varepsilon_{0}\varepsilon\langle\alpha|\int d^{3}\mathbf{r}|\hat{\mathbf{E}}(\mathbf{r})|^{2}|\alpha\rangle-\varepsilon_{0}\varepsilon\langle0|\int d^{3}\mathbf{r}|\hat{\mathbf{E}}(\mathbf{r})|^{2}|0\rangle$,
where $|\alpha\rangle$ is a state with a single photon in mode $\alpha$,
and $|0\rangle$ is the cavity vacuum. One obtains 
\begin{equation}
\hbar\omega_{\alpha}=2\varepsilon_{0}\varepsilon\int d\mathbf{r}|\mathbf{E}_{\alpha}(\mathbf{r})|^{2}\,.\label{eq:E_norm}
\end{equation}
With this normalization, Eq. \eqref{eq:G_jbg-1} reads
\begin{equation}
G_{\beta\gamma}^{j}=-i\frac{f\,M_{{\rm s}}}{8S\varepsilon}\sqrt{\omega_{\beta}\omega_{\gamma}}\xi\epsilon_{jmn}\frac{\int{\rm d}\mathbf{r}E_{\beta m}^{*}(\mathbf{r)}E_{\gamma n}(\mathbf{r})}{\sqrt{\int d\mathbf{r}|\mathbf{E}_{\beta}(\mathbf{r})|^{2}}\sqrt{\int d\mathbf{r}|\mathbf{E}_{\gamma}(\mathbf{r})|^{2}}}\,.\label{eq:G_norm}
\end{equation}

\begin{figure}
\begin{centering}
\includegraphics[width=0.6\textwidth]{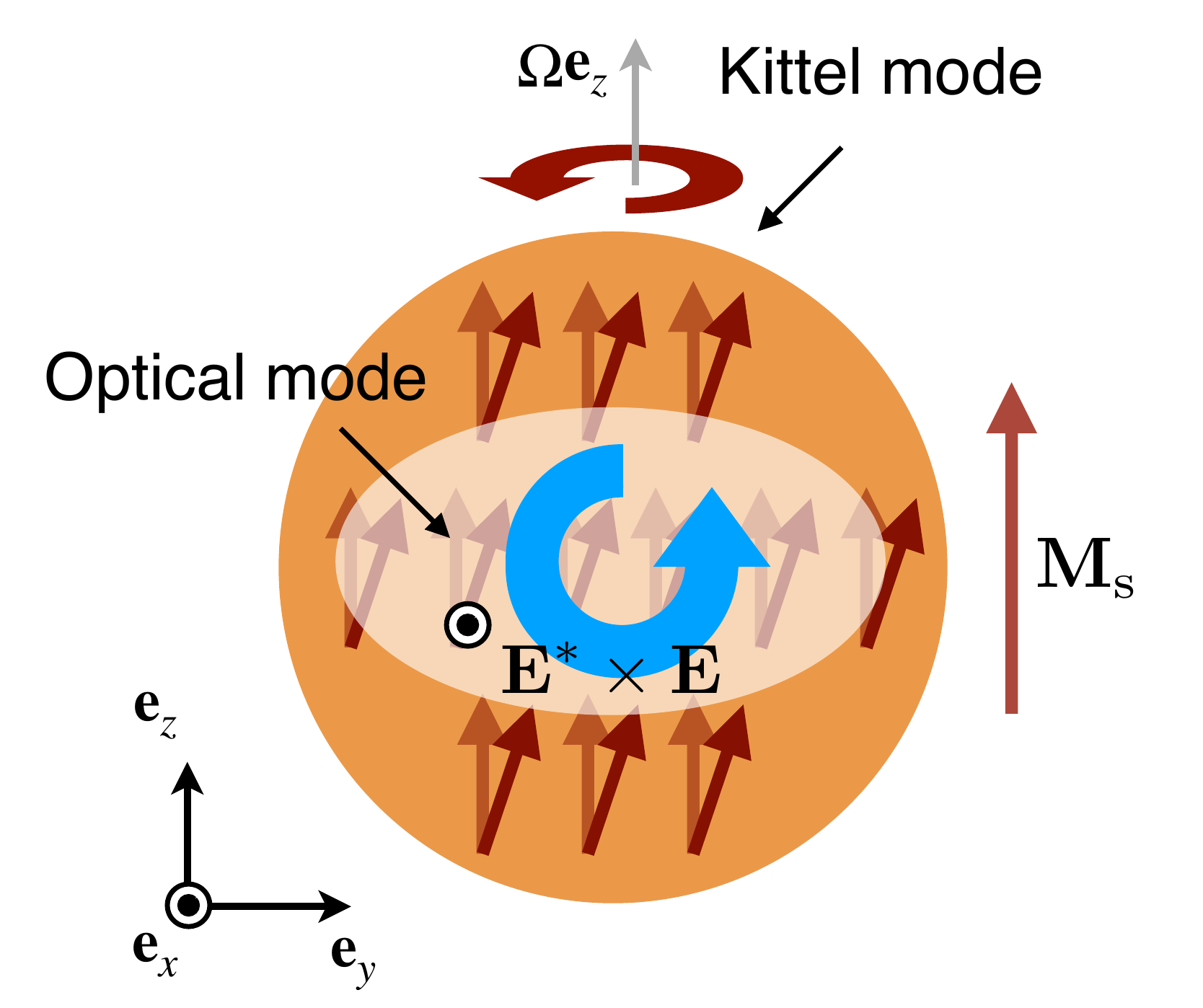}
\par\end{centering}
\caption{Schematic representation of the geometry for the calculation of the
optomagnonic coupling to the Kittel mode, Eqs. \eqref{eq:G_def} and
\eqref{eq:KittelHMO}. Adapted from Ref. \cite{kusminiskiyQuantumMagnetismSpin2019}.}

\label{Fig2}
\end{figure}
\begin{figure}
\begin{centering}
\includegraphics[width=0.8\textwidth]{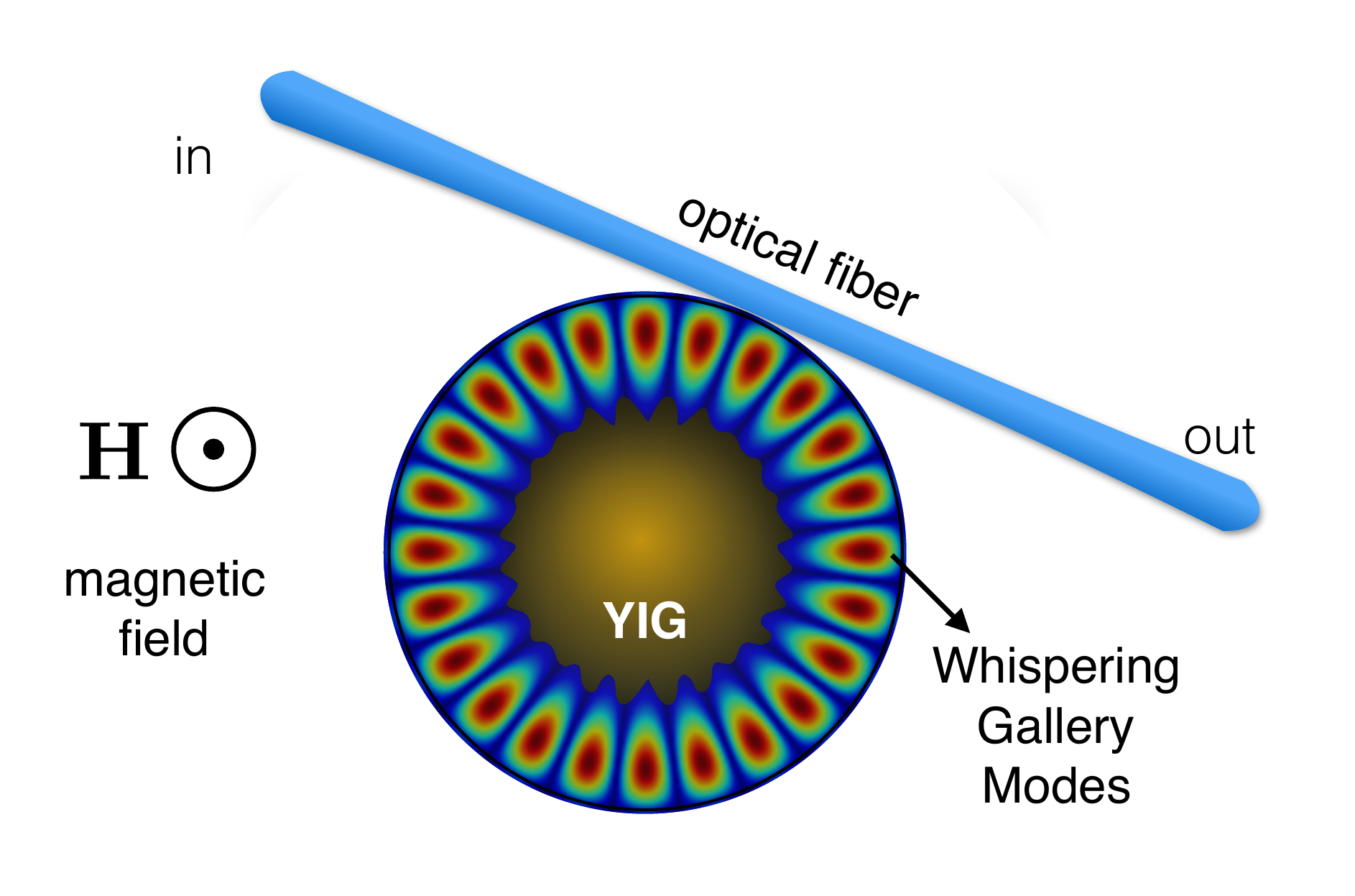}
\par\end{centering}
\caption{Sketch of a YIG sphere supporting optical WGM. Within the material,
photons and magnons interact via the optomagnonic interaction. The
frequency of the magnons can be controlled by an external magnetic
field $\mathbf{H}$ and the optical modes can be driven and probed
by an optical fiber, which couples evanescently to the WGM. }

\label{fig3}
\end{figure}

For processes involving a single optical mode, as we already pointed
out, some degree of circular polarization of the mode is necessary
for a finite coupling. Assuming an optical mode circularly polarized
in the $yz$ plane, the optical spin density is along the $x$ axis
and couples to the $x$ component of the spin operator, $\hat{S}_{x}$
(see Fig. \eqref{Fig2}). The relevant coupling matrix is therefore
$G_{\alpha\beta}^{x}$ which, in the considered geometry, is diagonal
in the circularly polarized basis for the optical fields $\mathbf{e}_{{\rm R}/{\rm L}}=\left(\mathbf{e}_{y}\mp i\mathbf{e}_{z}\right)/\sqrt{2}$,
as one can easily obtain from \eqref{eq:G_jbg-1}. We quantize the
optical field for simplicity in terms of plane waves
\begin{align}
\mathbf{E}(\mathbf{r},t)\rightarrow\mathbf{\hat{E}^{+}}(\mathbf{r},t) & =i\sum_{_{j}}\mathbf{e}_{j}\sqrt{\frac{\omega}{2\varepsilon V}}\hat{a}_{j}(t)e^{i\mathbf{k_{j}\cdot r}}\label{eq:E Q PW}\\
\mathbf{E}^{*}(\mathbf{r},t)\rightarrow\mathbf{\hat{E}^{-}}(\mathbf{r},t) & =-i\sum_{_{j}}\mathbf{e}_{j}\sqrt{\frac{\omega}{2\varepsilon V}}\hat{a}_{j}^{\dagger}(t)e^{-i\mathbf{k_{j}\cdot r}}\,,\nonumber 
\end{align}
where the $\pm$ superscripts follows the usual convention which indicates
the positive and negative frequency components of the optical field,
$\omega=\omega_{{\rm R}}=\omega_{{\rm L}}$, $j={\rm R},{\rm L}$
and $\mathbf{k}_{j}$ the corresponding wave vector. $V$ is the volume
of the optical cavity. Using these expressions, one can easily show
that Eq. \eqref{eq:G_norm} reduces to 
\begin{equation}
G=G_{{\rm LL}}^{x}=-G_{{\rm RR}}^{x}=\frac{1}{S}\frac{c\theta_{{\rm F}}}{4\sqrt{\varepsilon}}\xi\label{eq:G_def}
\end{equation}
where we have used Eq. \eqref{eq:theta F} to write $f$ in terms
of $\theta_{{\rm F}}$. The numerical factor $\xi\le1$ takes into
account the mode overlap of the electric field with the magnon mode and other geometric
factors. In current optomagnonic experiments, involving YIG spheres,
the optical modes in the cavity are actually whispering gallery modes
(WGM), see Fig. \eqref{fig3}. We will discuss a system with optical
whispering gallery modes in Sec. \eqref{sec:Optomagnonics-with-a},
for now these details are hidden in $\xi$. 

Considering Eqs. \eqref{eq:OM Ham coupling}, \eqref{eq:E Q PW},
and \eqref{eq:G_def}, the coupling Hamiltonian reads \cite{violakusminskiyCoupledSpinlightDynamics2016}
\begin{equation}
\hat{H}_{MO}=\hbar\hat{S}_{x}G\left(\hat{a}_{L}^{\dagger}\hat{a}_{L}-\hat{a}_{R}^{\dagger}\hat{a}_{R}\right)\,.\label{eq:KittelHMO}
\end{equation}
In the spin-wave approximation, for small oscillations of the macrospin
around its equilibrium position, we can replace the spin operator
$\hat{S}_{x}$ by a position operator $\hat{S}_{x}\rightarrow\sqrt{S/2}(\hat{m}+\hat{m}^{\dagger})$
using a Holstein-Primakoff transformation truncated to first order
in the bosonic operators (see Eq. \eqref{eq:Holstein Primakoff}).
In this limit,
\begin{equation}
\hat{H}_{MO}\approx\hbar\frac{1}{\sqrt{2S}}\frac{c\theta_{{\rm F}}}{4\sqrt{\varepsilon}}\xi\left(\hat{a}_{L}^{\dagger}\hat{a}_{L}-\hat{a}_{R}^{\dagger}\hat{a}_{R}\right)(\hat{m}+\hat{m}^{\dagger})\label{eq:H MO optomechanics}
\end{equation}
which is reminiscent of the Hamiltonian in the related field
of cavity optomechanics, where light couples to mechanical vibrations
by pressure forces. The coupling $g_{0}=G\sqrt{S/2}$ (given as an
angular frequency) is a measure of the single photon-magnon coupling
and equivalent to the vacuum coupling strength in optomechanics, where
$g_{0}$ is proportional to the zero-point motion of the mechanical
oscillator \cite{aspelmeyerCavityOptomechanics2014}. As we see from
the dependence on $1/\sqrt{S}$ in Eq. \eqref{eq:H MO optomechanics},
the single photon-magnon coupling is enhanced by small magnetic volumes. 

The material of choice for optomagnonic systems is the insulating
ferrimagnet YIG, due to the small losses both for the optics (absorption
coefficient $\alpha\sim0.069{\rm cm}^{-1}$ at $\lambda=1,2\,\mu{\rm m}$)
and for the magnon modes (Gilbert damping coefficient $\eta_{{\rm G}}\approx10^{-4}$)
and large Faraday rotation ($\theta_{{\rm F}}=240\,{\rm deg}/{\rm cm}$
at $\lambda=1,2\,\mu{\rm m}$). Note that although YIG is technically
a ferrimagnet, one magnetic sublattice is dominant and mostly behaves
like a ferromagnet. The optical cavity is formed by the magnetic material
itself, due to total internal reflection of light inside of the dielectric
material. At optical frequencies, the index of refraction of YIG is
$n=\sqrt{\varepsilon}\approx2.24$, which combined with its low absorption
makes it a reasonably good optical cavity if patterned appropriately.
Experiments so far have used YIG spheres, since they are commercially
available and relatively easy to polish into small sizes while preserving
the quality of the optical cavity. Sizes nevertheless remain still
too large, in the range of $100\mu{\rm m}$ radius. The YIG sphere supports optical modes in the way of whispering
gallery modes which can be accessed through a tapered fiber, see the
scheme of Fig. \eqref{fig3}. 

Assuming optimal mode-matching and a diffraction-limited volume of
YIG of $1\mu{\rm m}^{3}$, one obtains $G\sqrt{S/2}\approx0.1{\rm MHz}$
\cite{violakusminskiyCoupledSpinlightDynamics2016}, which would be
comparable to state of the art optomechanical systems \cite{maccabePhononicBandgapNanoacoustic2019}.
Current experimental setups are still far from this limit, due to
fabrication and design issues. Note for example that for the case
of a sphere, the Kittel mode is a bulk mode, whereas the WGMs live
near the surface, leading to a small overlap between the modes. Improving
the current values of the coupling is however highly desirable for
applications in the quantum regime.

\subsection{Magnetic textures}

One alternative to improve the value of the optomagnonic coupling
is to go beyond the Kittel mode, searching for modes that would be
better suited for mode matching with the optics. This is starting
to be explored both theoretically \cite{sharmaLightScatteringMagnons2017,grafCavityOptomagnonicsMagnetic2018a,sharmaOptimalModeMatching2019}
and experimentally \cite{haighSelectionRulesCavityenhanced2018a,osadaBrillouinLightScattering2018}.
This brings us to case (ii) from our two limiting cases, where one
allows for non-uniform ground states (also called \emph{magnetic textures})
and/or magnetic excitations with a spatial structure, and uses the
Holstein-Primakoff transformations to represent the excitations in
terms of bosonic operators $\hat{m}_{i}$, where $i$ is the lattice
site:

\begin{align}
\hat{S}_{i}^{+} & =\sqrt{2s}\sqrt{1-\frac{\hat{m}_{i}^{\dagger}\hat{m}_{i}}{2s}}\hat{m}_{i}\nonumber \\
\hat{S}_{i}^{-} & =\sqrt{2s}\hat{m}_{i}^{\dagger}\sqrt{1-\frac{\hat{m}_{i}^{\dagger}\hat{m}_{i}}{2s}}\nonumber \\
\hat{S}_{i}^{z} & =\left(s-\hat{m}_{i}^{\dagger}\hat{m}_{i}\right)\,.\label{eq:Holstein Primakoff}
\end{align}
In these, $s$ is the total spin per lattice site $i$, so that the
total spin is given by $S=Ns$ with $N$ the number of lattice sites,
and $\hat{S}_{i}^{\pm}=\hat{S}_{i}^{x}\pm i\hat{S}_{i}^{y}$ are the
spin ladder operators. Eqs. \eqref{eq:Holstein Primakoff} assume
a quantization axis along the $z$ direction. If the magnetic ground
state is textured, the quantization axis is local, defined by $\mathbf{e}_{z}(\mathbf{r})=\mathbf{M}_{{\rm s}}(\mathbf{r})/M_{{\rm s}}$.
The bosonic operators $\hat{m}_{i}$ fulfill the usual bosonic commutation
rules
\begin{align}
\left[\hat{m}_{i},\,\hat{m}_{j}^{\dagger}\right] & =\delta_{ij}\label{eq:comm ms}\\
\left[\hat{m}_{i},\,\hat{m}_{j}\right] & =\left[\hat{m}_{i}^{\dagger},\,\hat{m}_{j}^{\dagger}\right]=0\,.\nonumber 
\end{align}
The problem is simplified by cutting off the Holstein-Primakoff transformation
to first order in the bosonic operators,
\begin{align}
\hat{S}_{i}^{+} & \approx\sqrt{2S}\hat{m}_{i}\label{eq:ladder harmonic}\\
\hat{S}_{i}^{-} & \approx\sqrt{2S}\hat{m}_{i}^{\dagger}\nonumber \\
\hat{S}_{i}^{z} & \approx S\,,\nonumber 
\end{align}
and is therefore a linear approximation for the local spin operators,
which are treated as harmonic oscillators. The elementary magnetic
excitations are collective, since given a spin Hamiltonian (e.g. the
Heisenberg Hamiltonian), after performing the approximation Eq. \eqref{eq:ladder harmonic}
one still needs to bring the Hamiltonian to a diagonal form, so that
is is a sum of \emph{independent }harmonic oscillators (e.g. in the bulk by going
to Fourier space, $\hat{m}_{\mathbf{k}}$). These collective excitations
are denominated \emph{magnons}: essentially, one magnon is a ``flipped''
spin which is shared by the whole system. Higher-order terms in the
expansion can be included and represent magnon-magnon interactions.

The quantization of the coupling term Eq. \eqref{eq:MO energy} in
this case follows by writing the excitation $\delta\mathbf{M}(\mathbf{r},t)$
in terms of the magnon modes (by magnon modes we mean the bosonic
operators which diagonalize the magnetic Hamiltonian). It is convenient
to work in terms of the normalized magnetization $\delta\mathbf{m}(\mathbf{r},t)=\delta\mathbf{M}(\mathbf{r},t)/M_{{\rm s}}$.
For small deviations $|\delta\mathbf{m}|\ll1$ we can quantize the
spin wave in analogy to Eq. \eqref{eq:EfieldQuant} for the electric
fields, by the substitution 
\begin{equation}
\delta\mathbf{m}(\mathbf{r},t)\rightarrow\frac{1}{2}\sum_{\gamma}\left[\delta\mathbf{m}_{\gamma}(\mathbf{r})\hat{m}_{\gamma}+\delta\mathbf{m}_{\gamma}^{*}(\mathbf{r})\hat{m}_{\gamma}^{\dagger}\right]\,,\label{eq:delta_m}
\end{equation}
where $\hat{m}_{\gamma}^{(\dagger)}$ annihilates (creates) a magnon
in mode $\gamma$. The information on its spatial structure is contained
in the mode functions $\delta\mathbf{m}_{\gamma}(\mathbf{r})$. Together
with Eq. \eqref{eq:EfieldQuant}, from Eq. \eqref{eq:MO energy} and
using Eq. \eqref{eq:theta F} we obtain the optomagnonic coupling
Hamiltonian linearized in the spin fluctuations \cite{grafCavityOptomagnonicsMagnetic2018a}
\begin{equation}
\hat{H}_{MO}=\hbar\sum_{\alpha\beta\gamma}G_{\alpha\beta\gamma}\hat{a}_{\alpha}^{\dagger}\hat{a}_{\beta}\hat{m}_{\gamma}+ \hbar\sum_{\alpha\beta\gamma}G_{\alpha\beta\gamma}^{{\rm *}}\hat{a}_{\beta}^{\dagger}\hat{a}_{\alpha}\hat{m}_{\gamma}^{\dagger}\label{eq:linSHMO}
\end{equation}
where 
\begin{equation}
G_{\alpha\beta\gamma}^{{\rm }}=-i\frac{\theta_{{\rm F}}\lambda_{n}}{4\pi}\frac{\varepsilon_{0}\varepsilon}{2\hbar}\int{\rm d}\mathbf{r}\,\delta\mathbf{m}_{\gamma}(\mathbf{r})\cdot[\mathbf{E_{\alpha}^{*}}\left(\mathbf{r}\right)\times\mathbf{E}_{\beta}\left(\mathbf{r}\right)]\label{eq:coupling}
\end{equation}
is the optomagnonic coupling in terms of the Faraday rotation per
wavelength of the light in the material $\lambda_{n}=\lambda_{0}/n$,
with $\lambda_{0}$ the vacuum wavelength. For YIG one obtains $\theta_{{\rm F}}\lambda_{n}/2\pi\approx10^{-5}$.
Within the linear regime for the spins, Eq. \eqref{eq:m_norm} is
very general and allows to treat arbitrary geometries and modes, both
optical and magnetic. 

In Eq. \eqref{eq:coupling} one still needs to specify the normalization
of the modes. For the optical fields the
normalization was given in Eq. \eqref{eq:E_norm}. For the magnon
modes, we impose a total magnetization corresponding to one Bohr magneton
(times the corresponding gyromagnetic factor $g$) in the excitation.
This normalizes the coupling to one magnon. From the definition Eq.
\eqref{eq:delta_m}, this is equivalent to imposing \cite{grafCavityOptomagnonicsMagnetic2018a}
\begin{equation}
\frac{1}{4}\int{\rm d}\mathbf{r}\,|\delta\mathbf{m}_{\gamma}(\mathbf{r})|^{2}=\frac{g\mu_{{\rm B}}}{M_{{\rm s}}}\,.\label{eq:m_norm}
\end{equation}
The normalized coupling therefore reads
\begin{align}
G_{\alpha\beta\gamma}^{{\rm }}&=-i\frac{\theta_{{\rm F}}\lambda_{n}}{4\pi}\frac{1}{2}\sqrt{\frac{g\mu_{{\rm B}}}{M_{{\rm s}}}}\sqrt{\omega_{\beta}\omega_{\alpha}}\\ & \times\frac{\int{\rm d}\mathbf{r}\,\delta\mathbf{m}_{\gamma}(\mathbf{r})\cdot[\mathbf{E_{\alpha}^{*}}\left(\mathbf{r}\right)\times\mathbf{E}_{\beta}\left(\mathbf{r}\right)]}{\sqrt{\int{\rm d}\mathbf{r}\,|\delta\mathbf{m}_{\gamma}(\mathbf{r})|^{2}}\sqrt{\int d\mathbf{r}|\mathbf{E}_{\beta}(\mathbf{r})|^{2}}\sqrt{\int d\mathbf{r}|\mathbf{E}_{\alpha}(\mathbf{r})|^{2}}}\,.\nonumber\label{eq:G norm 1}
\end{align}
For the optical fields, it is common to define an effective mode volume
$V_{{\rm E}}^{\alpha}$ 
\begin{equation}
V_{{\rm E}}^{\alpha}=\frac{\int d^{3}\mathbf{r}|\mathbf{E}_{\alpha}(\mathbf{r})|^{2}}{\max\left\{ |\mathbf{E}_{\alpha}(\mathbf{r})|^{2}\right\} }\,,\label{eq:V E eff}
\end{equation}
which for a homogeneous electric field reduces simply to the volume
occupied by the field. Analogously, we can define an effective magnetic
volume 
\begin{equation}
V_{{\rm M}}^{\gamma}=\frac{\int d^{3}\mathbf{r}|\delta\mathbf{m}_{\gamma}(\mathbf{r})|^{2}}{\max\left\{ |\delta\mathbf{m}_{\gamma}(\mathbf{r})|^{2}\right\} }\,,\label{eq:V M eff}
\end{equation}
according to which 
\begin{equation}
G_{\alpha\beta\gamma}^{{\rm }}=-i\frac{\theta_{{\rm F}}\lambda_{n}}{4\pi}\frac{1}{2}\sqrt{\frac{g\mu_{{\rm B}}}{M_{{\rm s}}}}\frac{\sqrt{\omega_{\beta}\omega_{\alpha}}}{\sqrt{V_{{\rm M}}^{\gamma}V_{{\rm E}}^{\alpha}V_{{\rm E}}^{\beta}}}\frac{\int{\rm d}\mathbf{r}\,\delta\mathbf{m}_{\gamma}(\mathbf{r})\cdot[\mathbf{E_{\alpha}^{*}}\left(\mathbf{r}\right)\times\mathbf{E}_{\beta}\left(\mathbf{r}\right)]}{|\delta\mathbf{m}_{\gamma}(\mathbf{r_{{\rm m}}})||\mathbf{E}_{\alpha}(\mathbf{r_{{\rm m}}})||\mathbf{E}_{\beta}(\mathbf{r_{{\rm m}}})|}\,,\label{eq:G_norm_SW_full}
\end{equation}
where for simplicity of notation we have defined $\mathbf{r_{{\rm m}}}$
such that $|\mathbf{\delta\mathbf{m}_{\gamma}}(\mathbf{r_{{\rm m}}})|=\max\left\{ |\mathbf{\delta\mathbf{m}_{\gamma}}(\mathbf{r})|\right\} $
and analogously for $\mathbf{E_{\alpha}}$. 

From Eq. \eqref{eq:G_norm_SW_full} we see that the strength of the
coupling is cut off by the smallest volume in the integral factor.
Assuming similar effective mode volumes for the optical fields, $V_{{\rm E}}^{\alpha}\equiv V_{{\rm E}}^{\alpha}\approx V_{{\rm E}}^{\beta}$,
if $V_{{\rm M}}^{\gamma}\le V_{{\rm E}}^{\alpha}$ the coupling is
suppressed by a factor $\sqrt{V_{{\rm M}}^{\gamma}}/V_{{\rm E}}^{\alpha}$,
favoring small optical volumes. For $V_{{\rm M}}^{\gamma}\ge V_{{\rm E}}^{\alpha}$
instead, the coupling goes as $1/\sqrt{V_{{\rm M}}^{\gamma}}$, favoring
small magnetic volumes. Recalling that 

\begin{equation}
M_{{\rm s}}=\frac{Sg\mu_{B}}{V_{{\rm M}}^{\gamma}}\label{eq:Ms muB}
\end{equation}
 where $S=Ns$ is the total spin in the volume $V_{{\rm M}}^{\gamma}$
($N$ number of spins, $s$ spin value), we recover in this case the
behavior $\propto1/\sqrt{S}$ found above for the Kittel mode in the
spin-wave approximation. From these scaling arguments, we see that
small mode volumes and optimal mode matching are required for large
coupling. In Sec. \eqref{sec:Optomagnonics-with-a} we will use Eq.
\eqref{eq:G_norm_SW_full} to calculate the optomagnonic coupling
in a cavity system consisting of a micromagnetic disk.

\section{Total Hamiltonian\label{subsec:Total-Hamiltonian}}

In the previous section, we derived the Hamiltonian that governs the
coupling between optical photons and magnons in a cavity. In order
to study dynamical processes, we need the total Hamiltonian of the
system. Besides the coupling Hamiltonian, we need to include the free
Hamiltonian both for magnons and photons (the kinetic terms). The
cavity is an open system, which can be driven and is also subject
to dissipation, both for magnons and photons. We can include the driving
term in the Hamiltonian, the dissipation terms we will include at
the level of the equations of motion in Sec. \eqref{sec:Equations-of-Motion}. 

\subsection{Free Hamiltonian}

The total optomagnonic Hamiltonian $\hat{H}$ consists of the optomagnonic
coupling term, given by either Eqs. \eqref{eq:OM Ham coupling} and
\eqref{eq:G_jbg-1} or Eqs.\eqref{eq:linSHMO} and \eqref{eq:coupling},
plus the free photon Hamiltonian
\begin{equation}
\hat{H}_{{\rm ph}}=\hbar\sum_{\alpha}\omega_{\alpha}\hat{a}_{\alpha}^{\dagger}\hat{a}_{\alpha}\label{eq:H ph}
\end{equation}
and the free magnetic term $\hat{H}_{{\rm m}}$. For the Kittel mode
(case (i) from the previous section), this free term is simply the
Larmor precession of the macrospin $\hat{\mathbf{S}}$,
\begin{equation}
\hat{H}_{{\rm m}}^{{\rm K}}=-\hbar\Omega\hat{S}_{z}\,,\label{eq:freeKittel}
\end{equation}
where we have assumed an external magnetic field $\mathbf{B_{0}}$
applied along the \textbf{$\hat{\mathbf{e}}_{z}$ }axis and $\Omega$
is the free precession frequency of the macrospin, see Fig. \eqref{Fig2}.\textbf{
}We assume also that the ground state magnetization is saturated and
along \textbf{$\hat{\mathbf{e}}_{z}$.} The frequency $\Omega$ is
in general controlled by\textbf{ }$\mathbf{B_{0}}$. In the case of
a spherical magnet, due to the high symmetry of the system $\Omega$
is independent of the demagnetization fields \cite{kittelPhysicalTheoryFerromagnetic1949},
and given simply by 
\begin{equation}
\hbar\Omega_{{\rm sphere}}=|g\mu_{{\rm B}}B_{0}|\,.\label{eq:w sphere}
\end{equation}
For YIG, $g=2$ and the gyromagnetic factor equals that of the electron:
$|\gamma_{{\rm e}}|=|g\mu_{{\rm B}}/\hbar|=1.76\times10^{11}{\rm rad}/{\rm s}\cdot{\rm T}$.
Applied magnetic fields in the range of tens of ${\rm mT}$ therefore
lead to frequencies in the ${\rm GHz}$ range. For other geometries,
e.g. ellipsoids or thin films, $\Omega$ depends also on the demagnetization
fields which can be taken into account through demagnetization factors
\cite{osbornDemagnetizingFactorsGeneral1945}. 

For general magnon modes in the spin-wave approximation (case (ii)),
one writes the free magnetic term also as a sum of harmonic oscillators
\begin{equation}
\hat{H}_{{\rm m}}^{SW}=\hbar\sum_{\beta}\Omega_{\beta}\hat{m}_{\beta}^{\dagger}\hat{m}_{\beta}\label{eq:freeSW}
\end{equation}
where $\Omega_{\beta}$ is the dispersion of the $\beta$ magnon mode.
Part of the problem in confined geometries is finding the
magnon modes and corresponding dispersion, and, except for very simple
geometries, micromagnetic simulations must be employed. Note that
by using the Holstein-Primakoff expression for $\hat{S}_{z}$, c.f.
Eq. \eqref{eq:Holstein Primakoff}, Eq. \eqref{eq:freeKittel} reduces,
asides from a constant, to an expression like Eq. \eqref{eq:freeSW}. 

\subsection{Driving term}

The cavity system can be driven by an external laser. The magnon modes
can in principle also be driven by an external MW field, but we will
not consider this in the following. The driving term can be included
in the Hamiltonian as
\begin{equation}
\hat{H}_{{\rm {\rm D}}}=i\epsilon_{\alpha}(\hat{a}_{\alpha}e^{i\omega_{L}t}-\hat{a}_{\alpha}^{\dagger}e^{-i\omega_{L}t})\,,\label{eq:H drive}
\end{equation}
where $\alpha$ indicates the mode that is being driven, $\omega_{L}$
is the laser frequency and 
\begin{equation}
\epsilon_{\alpha}=\hbar\sqrt{\frac{2\kappa_{\alpha}\mathcal{P}_{\alpha}}{\hbar\omega_{L}}}\label{eq:powerEps}
\end{equation}
depends on the driving laser power $\mathcal{P}_{\alpha}$ and on
the cavity decay rate $\kappa_{\alpha}$ of the pumped mode due to
the coupling to the the driving channel, e.g. an optical fiber or
waveguide. It is common to work in a rotating frame at the laser frequency
$\omega_{L}$, so that the trivial time dependence $e^{i\omega_{L}t}$
is removed. This is achieved by the unitary transformation $\hat{U}=e^{-i\omega_{L}t\hat{a}_{\alpha}^{\dagger}\hat{a}_{\alpha}}$
under which the Hamiltonian transforms as $\hat{H}\rightarrow\hat{U}\hat{H}\hat{U}^{\dagger}-i\hbar\hat{U}\frac{\partial\hat{U}^{\dagger}}{\partial t}$.
In the rotating frame, for a single photon mode $\alpha$ one obtains
\begin{equation}
\hat{H}_{{\rm ph}}+\hat{H}_{{\rm {\rm D}}}\rightarrow-\hbar\Delta_{\alpha}\hat{a}_{\alpha}^{\dagger}\hat{a}_{\alpha}+i\epsilon_{\alpha}(\hat{a}_{\alpha}-\hat{a}_{\alpha}^{\dagger})\label{eq:RF_trafo}
\end{equation}
where $\Delta_{\alpha}=\omega_{{\rm L}}-\omega_{\alpha}$ is the detuning
of the driving laser frequency with respect to the resonance frequency
of the optical cavity for the $\alpha$-mode. The generalization to
multiple driven modes is straightforward. If $\Delta_{\alpha}>0$
($\Delta_{\alpha}<0$ ) the system is said to be blue (red) detuned.
In the literature, it is usual to write the Hamiltonian in the rotating
frame omitting the driving term (second term on the RHS of Eq. \eqref{eq:RF_trafo}).
In that case, the driving term is added at the level of the equations
of motion, together with the dissipation and fluctuation terms.

\subsection{Total Hamiltonian for the Kittel mode}

The total cavity optomagnonic Hamiltonian, in the rotating frame and
omitting the driving and dissipation terms, are given in the following
both for the Kittel mode (case (i)) and in the spin-wave approximation
(case (ii)). For the Kittel mode we choose a simplified model in which
the light is circularly polarized in the $yz$ plane giving rise to
the coupling in Eq. \eqref{eq:KittelHMO}. Hence
\begin{equation}
\hat{H}_{{\rm K}}=-\hbar\Delta\hat{a}^{\dagger}\hat{a}-\hbar\Omega\hat{S}_{z}+\hbar G\hat{S}_{x}\hat{a}^{\dagger}\hat{a}\label{eq:H_tot_K}
\end{equation}
with $G$ given by Eq. \eqref{eq:G_def}. Since in this simple case
the Hamiltonian is diagonal in the circularly polarized basis, right
and left handed modes are not coupled and we can restrict the Hamiltonian
to a single photon mode, which we denote with the operator $\hat{a}$
(compare with Eq. \eqref{eq:KittelHMO}). Note also that, as long
as we work with the Voigt geometry, we can always find a system of
coordinates such that the Hamiltonian can be expressed as in Eq. \eqref{eq:H_tot_K}.
The Hamiltonian in Eq. \eqref{eq:H_tot_K} seems deceptively simple,
since, as we will see below, it leads to rich nonlinear dynamics even
in the classical limit. The \emph{parametric}
coupling in the photon operators (the coupling is a two-photon process) gives rise to nonlinearities even in the spin-wave approximation. These are equivalent to the nonlinear behavior present in optomechanical systems \cite{aspelmeyerCavityOptomechanics2014}. Retaining the full macrospin dynamics introduces new nonlinear behavior unique to optomagnonic systems.

\subsection{Total Hamiltonian and linearization}

In the spin-wave approximation the total Hamiltonian reads
\begin{equation}
\hat{H}_{{\rm SW}}=-\hbar\sum_{\alpha}\Delta_{\alpha}\hat{a}_{\alpha}^{\dagger}\hat{a}_{\alpha}+\hbar\sum_{\beta}\Omega_{\beta}\hat{m}_{\beta}^{\dagger}\hat{m}_{\beta}+\hbar\sum_{\alpha\beta\gamma}G_{\alpha\beta\gamma}^{{\rm }}\hat{a}_{\alpha}^{\dagger}\hat{a}_{\beta}\hat{m}_{\gamma}+{\rm h.c.\,}\label{eq:H_tot_SW}
\end{equation}
with $G_{\alpha\beta\gamma}^{{\rm }}$ given in Eq. \eqref{eq:G_norm_SW_full}.
This Hamiltonian is a three-particle interacting Hamiltonian. Diagonalization
is possible by linearizing the optical fields around the steady state
solutions, 
\begin{eqnarray}
\hat{a}_{\alpha}= & \langle\hat{a}_{\alpha}\rangle+\delta\hat{a}_{\alpha}\label{eq:a fluct}
\end{eqnarray}
such that ${\rm d}\langle\hat{a}_{\alpha}\rangle/{\rm d}t=0$ and
all the dynamics is contained in the fluctuation fields $\delta\hat{a}_{\alpha}$.
The input laser power determines the average number of photons circulating
in the cavity in mode $\alpha$, $n_{\alpha}=|\langle\hat{a}_{\alpha}\rangle|^{2}$.
Considering terms up to linear order in the fluctuations $\delta\hat{a}_{\alpha}$,
Eq. \eqref{eq:H_tot_SW} reduces to a quadratic Hamiltonian
\begin{align}
\hat{H}_{{\rm fl}} & =-\hbar\sum_{\alpha}\Delta_{\alpha}\hat{a}_{\alpha}^{\dagger}\hat{a}_{\alpha}+\hbar\sum_{\beta}\Omega_{\beta}\hat{m}_{\beta}^{\dagger}\hat{m}_{\beta}\label{eq:linSAHMO}\\
 & +\hbar\sum_{\alpha\beta\gamma}G_{\alpha\beta\gamma}^{{\rm }}\left(\sqrt{n_{\alpha}}\delta\hat{a}_{\beta}\hat{m}_{\gamma}+\sqrt{n_{\beta}}\delta\hat{a}_{\alpha}^{\dagger}\hat{m}_{\gamma}\right)+{\rm h.c.}\,.\nonumber
\end{align}
This kind of Hamiltonian is well know from quantum optics and related
systems (see e.g. Refs. \cite{wallsQuantumOptics2008,aspelmeyerCavityOptomechanics2014}),
and it can be turned into a \emph{parametric amplifier} ($\delta\hat{a}_{\beta}\hat{b}_{\gamma}$
and $\hat{m}_{\gamma}^{\dagger}\delta\hat{a}_{\beta}^{\dagger}$ terms)
or a \emph{beam splitter} Hamiltonian ($\delta\hat{a}_{\beta}\hat{m}_{\gamma}$
and $\hat{m}_{\gamma}^{\dagger}\delta\hat{a}_{\beta}^{\dagger}$)
by tuning the external laser driving frequency. The combination 
\begin{equation}
G_{{\rm eff}}=\sqrt{n}G\label{eq:Geff}
\end{equation}
(with indices as appropriate) shows that the coupling $G$ is enhanced
by the square root of the number of photons trapped in the cavity,
and can in this way be controlled.

\chapter{Equations of Motion\label{sec:Equations-of-Motion}}

In this section we split again for simplicity the discussion into
the two cases (i) and (ii) detailed above. We will obtain the equations
of motion for the macrospin dynamics, only valid for the Kittel mode
but retaining the spin algebra and the full non-linearity of the problem,
and and the spin-wave approximation, where we restrict the problem
to the dynamics of coupled harmonic oscillators. 

\section{Heisenberg equations of motion }

The Heisenberg equation of motion for an operator $\hat{\mathbf{O}}$
evolving under a Hamiltonian $\hat{H}$ is given by
\begin{align}
\hbar\frac{d\hat{\mathbf{O}}}{dt} & =i[\hat{H},\hat{\mathbf{O}}]\,.\label{eq:heis EOM}
\end{align}
In the macrospin approximation using $\hat{H}_{{\rm K}}$ given in
Eq. \eqref{eq:H_tot_K} and imposing the commutation relations$\left[\hat{a},\hat{a}^{\dagger}\right]=1$,
$\left[\hat{S}_{i},\hat{S}_{j}\right]=i\epsilon_{ijk}\hat{S}_{k}$
one obtains the following coupled equations of motion for the macrospin
and the light field
\begin{align}
\dot{\hat{a}} & =-i\left(G\hat{S}_{x}-\Delta\right)\hat{a}\label{eq:EOM_c_S}\\
\dot{\mathbf{\hat{\mathbf{S}}}} & =\left(G\hat{a}^{\dagger}\hat{a}\mathbf{\,e}_{x}-\Omega\,\mathbf{e}_{z}\right)\times\hat{\mathbf{S}}\,.\nonumber 
\end{align}
In the spin wave approximation, considering a general multimode system
given by the Hamiltonian in Eq. \eqref{eq:H_tot_SW} we obtain
\begin{align}
\dot{\hat{a}}_{\alpha} & =i\Delta_{\alpha}\hat{a}_{\alpha}-i\sum_{\beta\gamma}G_{\alpha\beta\gamma}\hat{a}_{\beta}\hat{m}_{\gamma}-i\sum_{\beta\gamma}G_{\beta\alpha\gamma}^{*}\hat{a}_{\beta}\hat{m}_{\gamma}^{\dagger}\nonumber \\
\dot{\hat{m}}_{\gamma} & =-i\Omega_{\gamma}\hat{m}_{\gamma}-i\sum_{\alpha\beta}G_{\alpha\beta\gamma}^{*}\hat{a}_{\beta}^{\dagger}\hat{a}_{\alpha}\,,\label{eq:EOM_SW}
\end{align}
and analogously for $\hat{a}_{\alpha}^{\dagger}$ and $\hat{m}_{\gamma}^{\dagger}$.
Eqs. \eqref{eq:EOM_c_S} and \eqref{eq:EOM_SW} are written in a rotating
frame but do not contain either driving or dissipative terms, we will
include these below. For now, we note that for $G=0$ Eqs. \eqref{eq:EOM_c_S}
decouple into a simple harmonic oscillator for the optical field,
and a Larmor precession equation for the spin operator. In particular,
the equation of motion for the spin in this case is $\dot{\mathbf{\hat{\mathbf{S}}}}=-\Omega\,\mathbf{e}_{z}\times\hat{\mathbf{S}}$, which reduces to the well known Landau-Lifschitz equation of motion for the magnetization
by taking the classical expectation values and proper rescaling. Note
that the Landau-Lifschitz equation of motion is therefore semiclassical,
since it is derived as the classical limit of the Heisenberg equation
of motion. 

\section{Dissipative terms}

The dissipative rates in cavity systems are very important since
they determine how fast information in the system is lost to the environment.
A very important figure of merit in hybrid systems is the \emph{cooperativity}
$\mathcal{C}$, which is defined as the ratio of the effective coupling
strength (see Eq. \eqref{eq:Geff}) to the decay channels in the system,
in our case the photon decay rate $\kappa$ and the magnon decay rate
which we take as $\Gamma=\Omega\eta_{{\rm G}}$,
\begin{equation}
\mathcal{C}=4\frac{G_{{\rm eff}}^{2}}{\kappa\Gamma}\,.\label{eq:cooperativity}
\end{equation}
Quantum protocols require at least $\mathcal{C}>1$, so that information
transfer can occur before the information is lost. Coherent state
transfer between magnons and photons requires moreover that $\mathcal{C}/n_{{\rm tm}}>1$,
where $n_{{\rm tm}}$ is the number of thermal magnons (the photon
environment can be safely assumed to be at zero temperature for optical
photons, since $k_{{\rm B}}T\ll\omega_{{\rm ph}}$).

\subsection{Landau-Lifschitz-Gilbert equation of motion}

To recover the classical dynamics of the spin, it is still necessary
to include dissipation. This is done phenomenologically by adding
a Gilbert damping term to the Landau-Lifschitz equation
\begin{equation}
\dot{\mathbf{S}}=-\Omega\,\mathbf{e}_{z}\times\mathbf{S}+\frac{\eta_{{\rm G}}}{S}(\mathbf{\dot{S}}\times\mathbf{S})\label{eq:LLG}
\end{equation}
where $\eta_{{\rm G}}$ is a characteristic of the magnetic material
and denominated the Gilbert damping coefficient. For YIG, $\eta_{{\rm G}}\approx10^{-4}$,
which is very low when compared to other magnetic materials. The Eq.
\eqref{eq:LLG} for the classical macrospin is denominated the \emph{Landau-Lifschitz-Gilbert}
equation of motion. Note that the dissipative term damps the precession
of the spin, but does not alter the norm of the vector. 

\subsection{Coupling to an external bath for the photon field}

The optical cavity fields are also subject to dissipative processes
due to the interaction with an environment. This can be modeled by
a thermal bath of harmonic oscillators $\hat{b}_{k}$ of frequency
$\omega_{k}$ which couple linearly to the photonic field with strength
$g_{k}$

\begin{equation}
\hat{H}_{{\rm env}}=\hbar\omega_{a}\hat{a}^{\dagger}\hat{a}+\sum_{k}\hbar\omega_{k}\hat{b}_{k}^{\dagger}\hat{b}_{k}+\hbar\sum_{k}\left(g_{k}\hat{a}^{\dagger}\hat{b}_{k}+g_{k}^{*}\hat{b}_{k}^{\dagger}\hat{a}\right)\,.\label{eq:h env}
\end{equation}
The environment is then integrated out and its effect is taken into
account by a dissipation term in the equations of motion for the degree
of freedom of interest, in this case, the optical field $\hat{a}$
\cite{meystreElementsQuantumOptics2007}. In the Markov approximation,
this procedure results in a \emph{quantum Langevin} equation of motion
\begin{equation}
\dot{\hat{a}}(t)=-\frac{\kappa}{2}\hat{a}(t)+\hat{F}(t)\,,\label{eq:L_EOM_c}
\end{equation}
where we have already transformed to the rotating frame with frequency
$\omega_{a}$. The environment-induced dissipation is encoded in the
\emph{cavity decay rate} $\kappa$, 
\begin{equation}
\kappa=\pi\mathcal{D}(\omega_{a})\left|g(\omega_{a})\right|^{2}\label{eq:kappa def}
\end{equation}
where $\mathcal{D}(\omega_{a})$ is the \emph{density of states} (DOS)
of the bath, which in the Markov approximation can be evaluated at
the cavity resonance frequency. $\hat{F}(t)$ is the \emph{noise operator
} 
\begin{equation}
\hat{F}(t)=-i\sum_{k}g_{k}\hat{b}_{k}(0)e^{i\left(\omega_{a}-\omega_{k}\right)t}\,,\label{eq:F def}
\end{equation}
which represents the random ``quantum kicks'' of the environment on
the cavity mode. The expectation value of the noise operator for a
reservoir in thermal equilibrium is zero 
\begin{equation}
\langle\hat{F}(t)\rangle_{R}=0\label{eq:F vacuum}
\end{equation}
and it is related to the cavity decay rate $\kappa$ by the \emph{fluctuation-dissipation
theorem}
\begin{equation}
\kappa=\frac{1}{\bar{n}}\int_{-\infty}^{\infty}{\rm d}\tau\langle\hat{F}^{\dagger}(\tau)\hat{F}(0)\rangle_{R}\label{eq:FDT}
\end{equation}
with 
\begin{equation}
\bar{n}=\langle\hat{b}^{\dagger}(\Omega)\hat{b}(\Omega)\rangle_{R}=\frac{1}{e^{\hbar\beta\Omega}-1}\,,\label{eq:n th}
\end{equation}
where $\beta=1/k_{{\rm B}}T$ , $k_{{\rm B}}$ the Boltzmann constant
and $T$ the temperature of the bath. The vacuum noise correlator
is local in time
\begin{equation}
\langle0|\hat{F}(t')\hat{F}^{\dagger}(t'')|0\rangle_{R}=\kappa\delta\left(t'-t''\right)\,,\label{eq:noise corr}
\end{equation}
in accordance with the Markov approximation: the dynamics of the bath
is fast compared to that of the cavity, and it has no memory. In \emph{input-output
theory}, the noise operator is normalized to an operator $\hat{a}_{{\rm in}}=\hat{F}/\sqrt{\kappa}$
such that
\begin{equation}
\langle0|\hat{a}_{{\rm in}}(t')\hat{a}_{{\rm in}}^{\dagger}(t'')_{{\rm }}|0\rangle_{R}=\delta\left(t'-t''\right)\label{eq:a in corr}
\end{equation}
represents the input noise, and Eq. \eqref{eq:L_EOM_c} reads
\begin{equation}
\dot{\hat{a}}(t)=-\frac{\kappa}{2}\hat{a}(t)+\sqrt{\kappa}\hat{a}_{{\rm in}}(t)\label{eq:LEOMinout}
\end{equation}
If the system is driven, the expectation value $<\hat{a}_{{\rm in}}(t)>$
is taken over a coherent state and it is finite, whereas for the fluctuations
$<\delta\hat{a}_{{\rm in}}(t)>=0$. The decay rate is in general
added to the equations of motion (e.g. in Eq. \eqref{eq:EOM_c_S})
as a phenomenological parameter, since the microscopic details of
the bath are usually not known. There can be several decay channels
for the photons in a cavity, due for example to scattering with phonons
or leaky mirrors. However, there are also ``wanted'' decay channels,
those which allow probing the cavity (for example, an external optical
fiber coupled to the cavity). Sometimes is it useful to split the
total decay rate into the unwanted losses $\kappa_{0}$ and the losses
associated with the input-output channel $\kappa_{{\rm in}}$, which
contain information about the state of the cavity system and can to
a certain degree be tuned. The total decay rate in this case (for
a high quality cavity) is simply the sum of the two contributions
$\kappa=\kappa_{{\rm in}}+\kappa_{0}$, and the quantum Langevin equation
reads
\begin{equation}
\dot{\hat{a}}(t)=-\frac{\kappa}{2}\hat{a}(t)+\sqrt{\kappa_{{\rm in}}}\hat{a}_{{\rm in}}(t)++\sqrt{\kappa_{{\rm 0}}}\hat{d}_{{\rm 0}}(t)\,,\label{eq:QLE a}
\end{equation}
where $\hat{d}_{{\rm 0}}$ is a noise operator associated with the
unwanted losses, with zero expectation value $\langle\hat{d}_{{\rm 0}}\rangle$.
It is sometimes customary to define the dimensionless parameter $\eta=\kappa_{0}/\kappa$
as the ratio of useful loss to total loss.

The procedure of ``integrating out'' in order to obtain an effective
equation of motion for the degrees of freedom of interest is quite
general. The Markov approximation is widely used for systems in contact
with a thermal reservoir, since its assumptions are in this case well
justified. The Langevin equation is an example of an equation of motion
which includes \emph{backaction}. In particular, the dissipative term
is the first correction to the instantaneous coupled system-bath dynamics,
due to the retardation in the response of the bath to a change in
the system of interest. In general the instantaneous response also
causes an energy shift in the frequency of the cavity field, which
we have ignored in the previous discussion. 

\section{\label{sec:Coupled-Equations-of}Light-induced dynamics of a classical
macrospin}

We now turn to the classical dynamics of the coupled photon-spin system,
following Ref. \cite{violakusminskiyCoupledSpinlightDynamics2016}.
We therefore replace the operators by their classical expectation
values $a=\langle\hat{a}\rangle$ and $\mathbf{S}=\langle\hat{\mathbf{S}}\rangle$,
\begin{eqnarray}
\dot{a} & = & -i\left(GS_{x}-\Delta\right)a-\frac{\kappa}{2}\left(a-\alpha_{{\rm max}}\right)\label{eq:cl_EOM_c}\\
\dot{\mathbf{S}} & = & \left(Ga^{*}a\mathbf{\,e}_{x}-\Omega\,\mathbf{e}_{z}\right)\times\mathbf{S}+\frac{\eta_{{\rm G}}}{S}(\mathbf{\dot{S}}\times\mathbf{S})\,,\label{eq:cl_EOM_S}
\end{eqnarray}
and ignore the fluctuations. In Eq. \eqref{eq:cl_EOM_c} we included
the driving laser amplitude $\alpha_{{\rm max}}$ for the optical
mode. This corresponds to the steady state value of $a$ for the uncoupled
system at zero detuning. As we anticipated, this is a highly nonlinear
system of equations and the full dynamics can be solved only numerically.
Analytical progress is however possible in certain limits. In particular,
in the \emph{fast cavity limit} we can follow the previous example
for a system in contact with a fast environment. In this case, however,
we integrate out the \emph{cavity} field, in order to obtain an effective
equation of motion for the macrospin. The term ``fast cavity'' refers
to the limit in which the photons in the cavity decay very rapidly
compared to the dynamics of the macrospin, so that a photon in the
cavity ``sees'' mostly a static spin. The fast cavity condition in
this case is given by $G\dot{S}_{x}\ll\kappa^{2}$. We will find that
the light field is responsible for extra dissipation and a frequency
shift for the spin precession. 

In order to find the effective equation of motion for the macrospin in this
limit, we expand the photon field $a(t)$ in powers of $\dot{S}_{x}$,
\begin{equation}
a(t)=a_{0}(t)+a_{1}(t)+\ldots\label{eq:c_exp}
\end{equation}
where the subscript indicates the order in the expansion in $\dot{S}_{x}$.
$a_{0}(t)$ therefore corresponds to the instantaneous equilibrium.
From Eq. \eqref{eq:cl_EOM_c}, imposing $\dot{a}(t)=0$ one finds
\begin{equation}
a_{0}(t)=\frac{\kappa}{2}\alpha_{{\rm max}}\frac{1}{\frac{\kappa}{2}-i\left(\Delta-GS_{x}(t)\right)}\,.\label{eq:a0}
\end{equation}
Inserting Eqs. \eqref{eq:c_exp} and \eqref{eq:a0} into Eq. \eqref{eq:cl_EOM_c}
and keeping terms to first order in $\dot{S}_{x}$ we obtain the correction
$a_{1}$ 
\begin{equation}
a_{1}(t)=-\frac{1}{\frac{\kappa}{2}-i\left(\Delta-GS_{x}\right)}\frac{\partial a_{0}}{\partial S_{x}}\dot{S}_{x}\,,\label{eq:a1}
\end{equation}
where we have used that $\dot{a}_{1}$ includes, by definition, only terms up to first order in $\dot{S}_{x}$.
Finally, replacing $|a|^{2}\approx|a_{0}|^{2}+a_{1}^{*}a_{0}+a_{0}^{*}a_{1}$
in Eq.~\eqref{eq:cl_EOM_S} and keeping again terms only up to first
order in $\dot{S}_{x}$, we obtain and effective equation of motion
for the classical macrospin $\mathbf{S}$ 
\begin{equation}
\dot{\mathbf{S}}=\mathbf{B}_{{\rm eff}}\times\mathbf{S}+\frac{\eta_{{\rm opt}}}{S}(\dot{S}_{x}\,\mathbf{e}_{x}\times\mathbf{S})+\frac{\eta_{{\rm G}}}{S}(\mathbf{\dot{S}}\times\mathbf{S})\,,\label{eq:EOM_eff}
\end{equation}
with $\mathbf{B}_{{\rm eff}}=-\Omega\mathbf{e}_{z}+\mathbf{B}_{{\rm opt}}$.
Both $\mathbf{B}_{{\rm opt}}$ and $\eta_{{\rm opt}}$ are light induced
and depend implicitly on time through $S_{x}(t)$. The quantity 
\begin{equation}
\mathbf{B}_{{\rm opt}}(S_{x})=G|a_{0}(S_{x})|^{2}\,\mathbf{e}_{x}\label{eq:Bopt}
\end{equation}
is the instantaneous response of the light field and acts as an optically
induced magnetic field, consequently giving rise to a frequency shift
for the precession of the spin. The second term in the RHS of Eq.
\eqref{eq:EOM_eff} is due to retardation effects, and it is reminiscent
of Gilbert damping, albeit with spin-velocity component only along
$\mathbf{\mathbf{e}}_{x}$ due to the geometry we chose, where the
spin of light lies along the $\mathbf{e}_{x}$ axis. 

The optically induced field $\mathbf{B}_{{\rm opt}}$ and the dissipation
coefficient $\eta_{{\rm opt}}$ are highly non-linear functions of
the spin coordinate $S_{x}(t)$,
\begin{eqnarray}
\mathbf{B}_{{\rm opt}} & = & \frac{G}{[(\frac{\kappa}{2})^{2}+(\Delta-GS_{x})^{2}]}\left(\frac{\kappa}{2}\alpha_{{\rm max}}\right)^{2}\mathbf{e}_{x}\label{eq:B_ind}\\
\eta_{{\rm opt}} & = & -2G\kappa S\,|\mathbf{B}_{{\rm opt}}|\,\frac{(\Delta-GS_{x})}{[(\frac{\kappa}{2})^{2}+(\Delta-GS_{x})^{2}]^{2}}\,,\label{eq:G_ind}
\end{eqnarray}
and can be tuned externally by the laser drive, both by the power
and by the detuning. The strength of the induced field is controlled
by a combination of the input power and the decay rate of the photons
in the cavity. The detuning however has a qualitative effect. In particular
when the condition $\Delta>G|S_{x}|$ is fulfilled, the optically
induced dissipation is \emph{negative}, leading to instabilities of
the original stable equilibrium points when it dominates over the
Gilbert damping $\eta_{{\rm G}}$. The instabilities are a consequence
of driving the system blue-detuned, which pumps energy into the spin
system. This can be seen by studying the stability of the north pole
(which is the stable solution without driving) once the driving laser is applied.
From Eq. \eqref{eq:EOM_eff} assuming $\eta_{G}\,\ll\eta_{{\rm opt}}$
(this condition can be easily achieved e.g. for YIG), we can obtain
an equation of motion for $S_{x}$. Setting $S_{z}=S$, 

\begin{equation}
\ddot{S}_{x}=-\Omega SB_{{\rm opt}}-\Omega^{2}S_{x}-\eta_{{\rm opt}}\Omega\dot{S}_{x}\,,\label{eq:S linearize}
\end{equation}
we consider small deviations $\delta S_{x}$ of $S_{x}$ from the
equilibrium position that satisfies $S_{x}^{0}=-SB_{{\rm opt}}/\Omega$,
where $B_{{\rm opt}}$ in Eq. \eqref{eq:Bopt} is evaluated at $S_{x}^{0}$.
To linear order we obtain 
\begin{align}
\ddot{\delta S}_{x} & =-\Omega\left(\Omega+S\frac{\partial B_{{\rm opt}}}{\partial S_{x}}\right)\delta S_{x}\label{eq:aux 1}\\
 & +2GS\kappa\Omega B_{{\rm opt}}\frac{(\Delta+GSB_{{\rm opt}}/\Omega)}{\left[\left(\kappa/\Omega\right)^{2}+(\Delta+GSB_{{\rm opt}}/\Omega)^{2}\right]^{2}}\dot{\delta S}_{x}\,.\nonumber 
\end{align}
Therefore the effective dissipation coefficient is in this case
\begin{equation}
\eta_{{\rm opt}}\approx-2GS\kappa B_{{\rm opt}}\frac{(\Delta+GSB_{{\rm opt}}/\Omega)}{\left[\left(\kappa/\Omega\right)^{2}+(\Delta+GSB_{{\rm opt}}/\Omega)^{2}\right]^{2}}\,,\label{eq:nopt eff}
\end{equation}
which is always negative for blue detuning. Comparing with Eq. \eqref{eq:aux 1}, we see that the solutions near
the north pole are unstable for $\Delta>0$. 

The runaway solutions can fall either into a new static equilibrium
point at more or less the opposite pole on the Bloch sphere (aligned
with the equilibrium value of $\mathbf{B}_{{\rm eff}}=-\Omega\mathbf{e}_{z}+\mathbf{B}_{{\rm opt}}$, but $|\mathbf{B}_{{\rm opt}}|\ll\Omega$) resulting in an effective switching of the
magnetization. The other possibility is to fall into a limit cycle
attractor, where the solution is a periodic motion of the spin on
the Bloch sphere. Which attractor is selected can be determined by
analyzing instead the stability near the south pole, $S_{z}=-S$.
In this case $S_{x}^{0}=SB_{{\rm opt}}/\Omega$ and 
\begin{align}
\ddot{\delta S}_{x} & -\Omega\left(\Omega-S\frac{\partial B_{{\rm opt}}}{\partial S_{x}}\right)\delta S_{x}\label{eq:aux 2}\\
 & -2GS\kappa\Omega B_{{\rm opt}}\frac{(\Delta-GSB_{{\rm opt}}/\Omega)}{\left[\left(\kappa/\Omega\right)^{2}+(\Delta-GSB_{{\rm opt}}/\Omega)^{2}\right]^{2}}\dot{\delta S}_{x}\,.\nonumber 
\end{align}
Therefore for $\Delta>GSB_{{\rm opt}}/\Omega$, the effective $\eta_{{\rm opt}}$
in this case is positive and the solution is a stable \emph{fixed} point resulting
in magnetization switching, which can be seen as a population inversion
driven by a blue detuned laser. In the opposite case ($\Delta<GSB_{{\rm opt}}/\Omega$),
$\eta_{{\rm opt}}<0$ and there are runaway solutions. These two instabilities
at the north and south pole are indicative of a limit cycle. In general,
the limit cycles behavior is due to the change of sign of the dissipation
function on the Bloch sphere (note that the condition $\Delta>G|S_{x}|$
can be fulfilled $\forall S_{x}$ or for just a region on the Bloch
sphere, depending on the magnitude of the detuning), analogous to
the Van der Pol oscillator dynamics. These self-sustained oscillations
can be the working principle for magnon lasing \cite{violakusminskiyCoupledSpinlightDynamics2016}. 

\begin{figure}[t]
\begin{centering}
\includegraphics[width=0.8\textwidth]{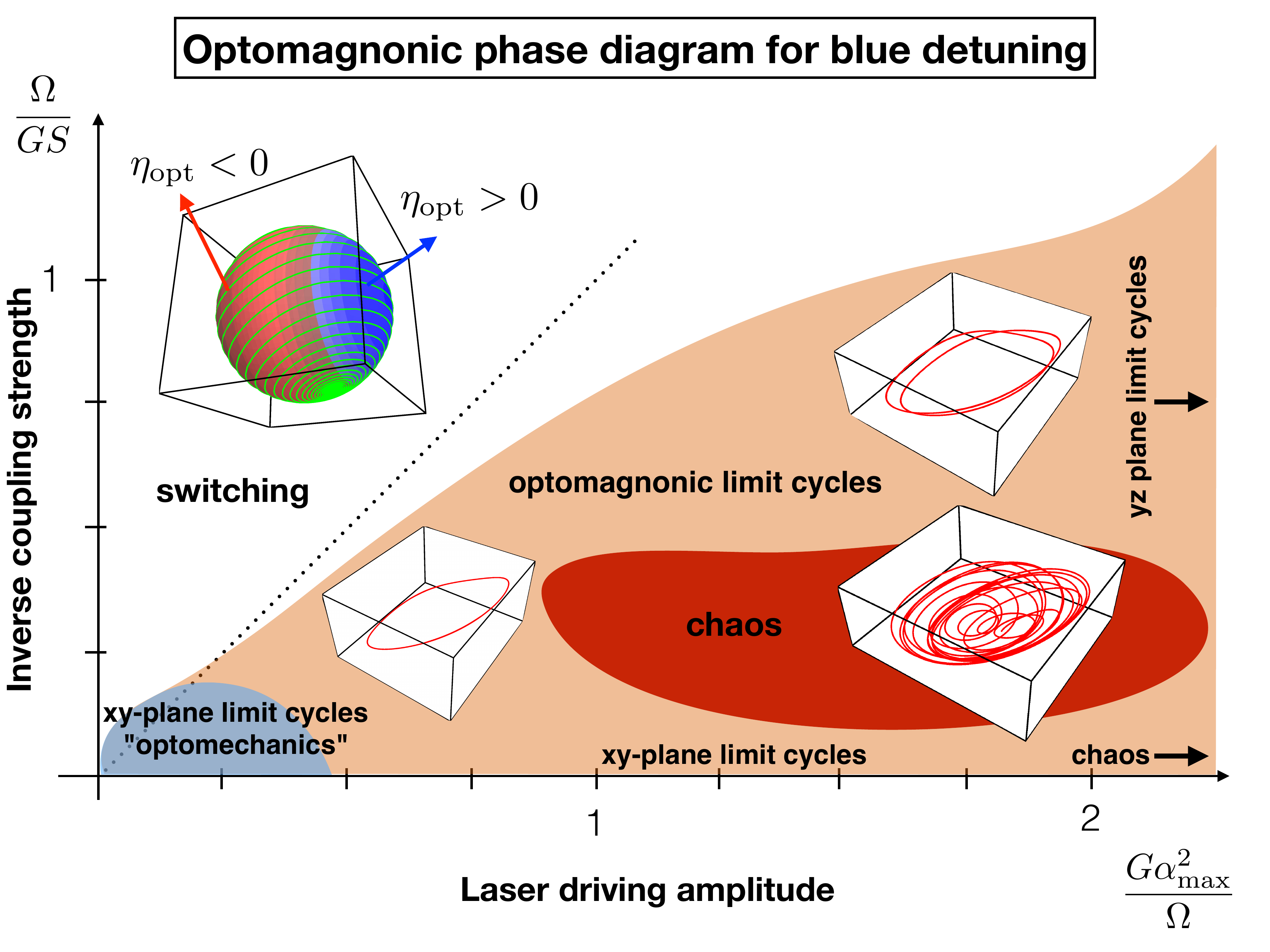}
\par\end{centering}
\caption{Qualitative phase diagram for the classical nonlinear behavior of
a blue detuned optomagnonic system as a function of the laser driving
strength and inverse optomagnonic coupling. The axis have been rescaled
into dimensionless quantities. The Bloch sphere in the switching region
shows an example of the trajectory of the macrospin (in green) to
the fixed point near the south pole, together with the sign of the
optically induced dissipation coefficient $\eta_{{\rm opt}}$. The
white boxes in the limit cycle and chaos region show the trajectory
of the macrospin (in red) once the corresponding attractor has been
reached, showing a single and double-period limit cycle and an example
of chaotic dynamics. Adapted from Ref. \cite{violakusminskiyCoupledSpinlightDynamics2016}.}

\label{Fig5}
\end{figure}

Beyond the fast cavity limit, Eqs. \eqref{eq:cl_EOM_c} and \eqref{eq:cl_EOM_S}
can be solved numerically. Besides magnetization switching and limit
cycles, the full dynamics of the system can be driven into a chaotic
regime, reached by period doubling as the power of the laser is increased.
The chaotic regime requires sideband resolution ($\Omega>\kappa$),
and strong coupling $G$ or equivalently a high density of circulating
photons in the cavity, determined by the laser power. The estimated
minimum values for attaining chaos seem to be outside of the current
capabilities with YIG, see Ref. \cite{violakusminskiyCoupledSpinlightDynamics2016}
for details. Whereas magnetization switching and self oscillations
can be attained at more moderate values of the parameters, other dissipation
processes not considered here (such as three and four-magnon scattering
processes) could hinder their realization in solid state systems \cite{clogstonFerromagneticResonanceLine1956}.
These regimes can however be attained in cavity cold atom systems
\cite{kohlerCavityAssistedMeasurementCoherent2017}.

In the limit of small oscillations of the macrospin, we can fix $S_{z}=S$
($\dot{S}_{z}=0$) and the remaining dynamical variables of the spin
$S_{x}$ and $S_{y}$ behave as the conjugate coordinates of a harmonic
oscillator. This is the classical limit of the Holstein-Primakoff
approximation and it is valid as long as $S_{x},S_{y}\ll S$. In this
limit (neglecting the Gilbert damping) we obtain 
\begin{align}
\dot{S}_{x} & =\Omega S_{y}\label{eq:simple S EOM}\\
\dot{S}_{y} & =-GS|c|^{2}-\Omega S_{x}\nonumber 
\end{align}
and hence
\begin{align}
\ddot{S}_{x} & =-GS\Omega|c|^{2}-\Omega^{2}S_{x}\label{eq:S HO}
\end{align}
with $c(t)$ given by Eq. \eqref{eq:cl_EOM_c}. The nonlinear dynamics
in this case reduces to the one known for optomechanical systems in
the classical limit \cite{marquardtDynamicalMultistabilityInduced2006}.
Fig. \eqref{Fig5} presents a qualitative, schematic phase diagram for
the optomagnonic system in the blue detuned regime as a function of
laser power and coupling strength, highlighting the different possible
nonlinear regimes. 

\chapter{Optomagnonics with a magnetic vortex\label{sec:Optomagnonics-with-a}}

In this section we come back to the issue of calculating the optomagnonic
coupling in the presence of a smooth magnetic textures and structured
optical modes, as given by Eq. \eqref{eq:G_norm_SW_full}. By smooth
magnetic textures we refer to a magnetization profile that can be
represented by a continuous vector field with constant length, given
in our case by $\mathbf{M}(\mathbf{r})=M_{{\rm s}}\mathbf{m}(r)$.
As an example, we will consider a cavity system with cylindrical geometry.
The optical cavity is a dielectric microdisk which can also host magnetic
modes. Due to the size and geometry, the magnetic ground state is a
vortex. We will calculate explicitly the optomagnonic coupling of
whispering gallery modes in the disk, to a magnetic excitation localized
at the vortex core. In this section we follow Ref. \cite{grafCavityOptomagnonicsMagnetic2018a}.

\subsection{Magnetic vortex}

\begin{figure}
\begin{centering}
\includegraphics[width=0.6\textwidth]{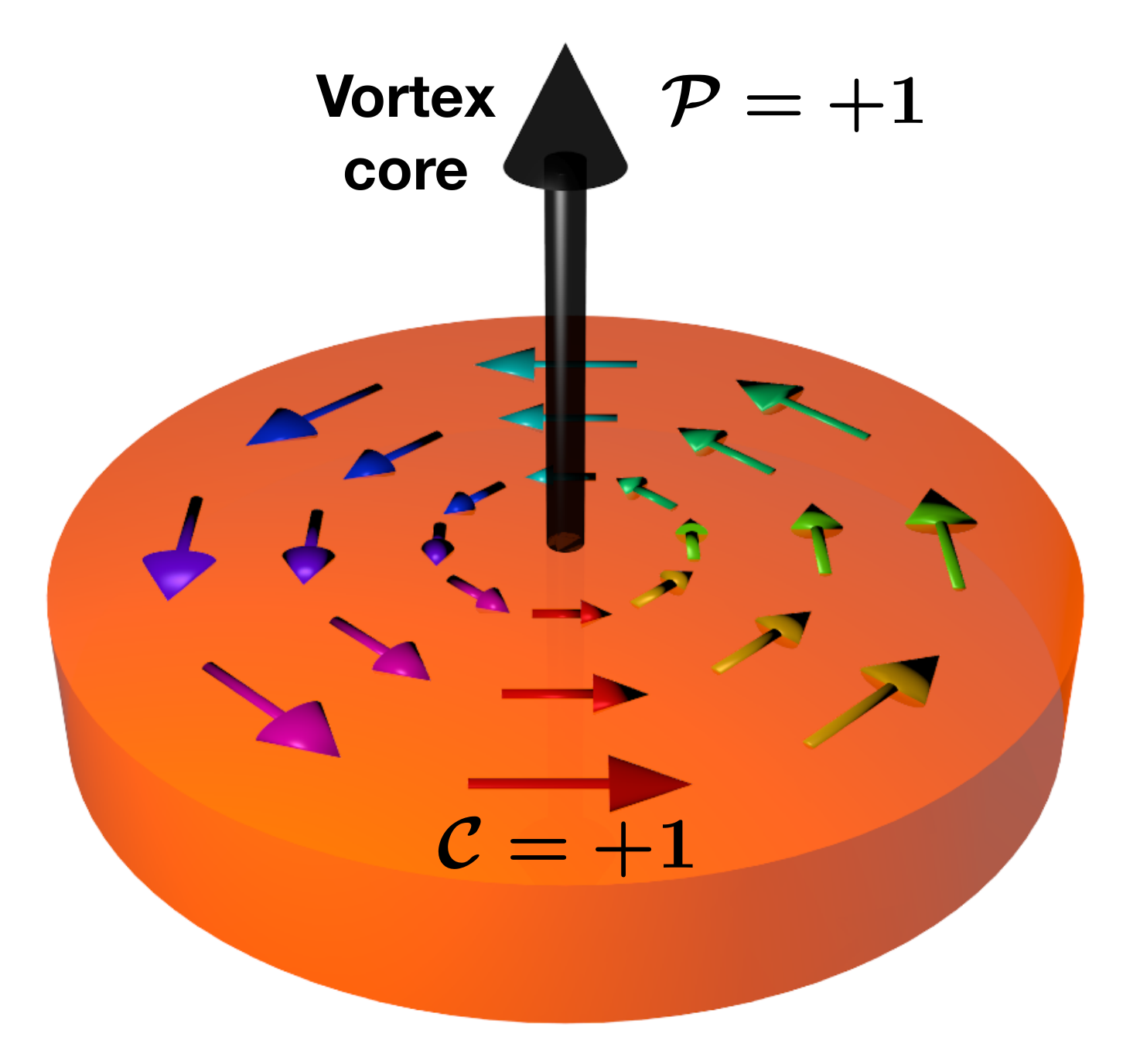}
\par\end{centering}
\caption{Magnetic vortex in a microdisk. The spins curl in the plane of the
disk and at the core of the vortex point out of the plane. Adapted
from Ref. \cite{grafCavityOptomagnonicsMagnetic2018a}.}

\label{Fig55}
\end{figure}

A paradigmatic example of a magnetic texture is a magnetic vortex,
which is the stable ground state configuration in magnetic disks with
radial dimensions in the $\mu{\rm m}$ range. The vortex forms due
to a competition between demagnetizing fields, which tend to align
with the surfaces of the disk to avoid the formation of sources of
stray fields, and the exchange interaction, which tends to align the
spins among themselves. The demagnetizing fields are determined by
the magnetostatic Poisson equation and have their origin in the dipolar
interactions, and are therefore weak but long ranged. The exchange
interaction, whose physical origin is the Coulomb interaction together
with the Pauli principle of exclusion, is instead strong but short
ranged. The effects of these fields become comparable at the microscale,
leading to ordered flux closure configurations that, in the case of
a cylindrical geometry, take the form of a vortex. In a thin disk
(for heights comparable to the exchange length $l_{{\rm ex}}$) the
spins are mostly in-plane and curl around the center of the disk.
The core of this vortex is situated at the center of the disk, and
it consists of spins pointing out of the plane \cite{guslienkoMagneticVortexState2008}.
Besides being ubiquitous, vortices are interesting since they are
topological objects with two independent degrees of freedom, the chirality
$\mathcal{C}=\pm1$ (the spins can curl clockwise or anti-clockwise)
and the parity $\mathcal{P}=\pm1$ , indicating if the spins at the
vortex core point up or down, see Fig. \eqref{Fig55}. Manipulating
these degrees of freedom can give rise to new forms of storing and
processing information with magnetic systems \cite{pigeauOptimalControlVortexcore2011}.

The vortex in the thin disk can be parametrized as
\begin{align}
\mathbf{m}(\vec{\rho}) & =\mathbf{e}_{\varphi} & {\rm for}\,\rho\ge b\label{eq:m_of_rho}\\
 & =\frac{1}{\rho^{2}+b^{2}}\left(\begin{array}{c}
-2by\\
2bx\\
\left(b^{2}-\rho^{2}\right)
\end{array}\right) & {\rm for}\,\rho\le b
\end{align}
where $b$ is an effective core radius (of the order of a few $l_{{\rm ex}}$), we
have assumed $,\mathcal{C}=\mathcal{P}=1$ and used cylindrical coordinates
with origin at the center of the vortex core, $\mathbf{e}_{\varphi}=(\cos\varphi,\sin\varphi,0)$. 

The lowest energy magnon mode in this system is a translational mode
of the vortex core, which can be shown to be a circular motion. This is due
to an effective gyrotropic force proportional to the topological charge
of the vortex, which effectively acts on the vortex core as a magnetic field acts on a charged particle
\cite{thieleSteadyStateMotionMagnetic1973}. This mode is denominated
gyrotropic and is usually in the range of hundreds of ${\rm MHz}$.
It gives rise to a time-dependent magnetization that can be approximated
as 
\begin{equation}
\mathbf{m_{{\rm ex}}}(\mathbf{r},t)=\mathbf{m}(\mathbf{r}-r_{c}(t))\approx\mathbf{m}(\mathbf{r})-\left(r_{c}(t)\cdot\nabla\right)\mathbf{m}(\mathbf{r})\,,\label{eq:m vortex app}
\end{equation}
where $\mathbf{r}_{c}(t)=r_{c}\left[\cos(\omega_{g}t)\mathbf{e}_{x}+\sin(\omega_{g}t)\mathbf{e}_{y}\right]$
parametrizes the time-dependent position of the vortex core measured
from its equilibrium position (we have ignored the damping of the
mode). Therefore we can write 
\begin{equation}
\delta\mathbf{m}(\mathbf{r},t)=-\left(r_{c}(t)\cdot\nabla\right)\mathbf{m}(\mathbf{r})\,.\label{eq:delta_m_of_r}
\end{equation}
From Eqs. \eqref{eq:m_of_rho} and \eqref{eq:delta_m_of_r} one can
obtain analytical expressions for the gyrotropic mode profile $\delta\mathbf{m}(\mathbf{r})$.
Together with the normalization prescription Eq. \eqref{eq:m_norm}
one obtains the profile of the mode normalized to one magnon \cite{grafCavityOptomagnonicsMagnetic2018a}.

\subsection{Optomagnonic coupling for the gyrotropic mode}

\begin{figure}
\begin{centering}
\includegraphics[width=0.8\textwidth]{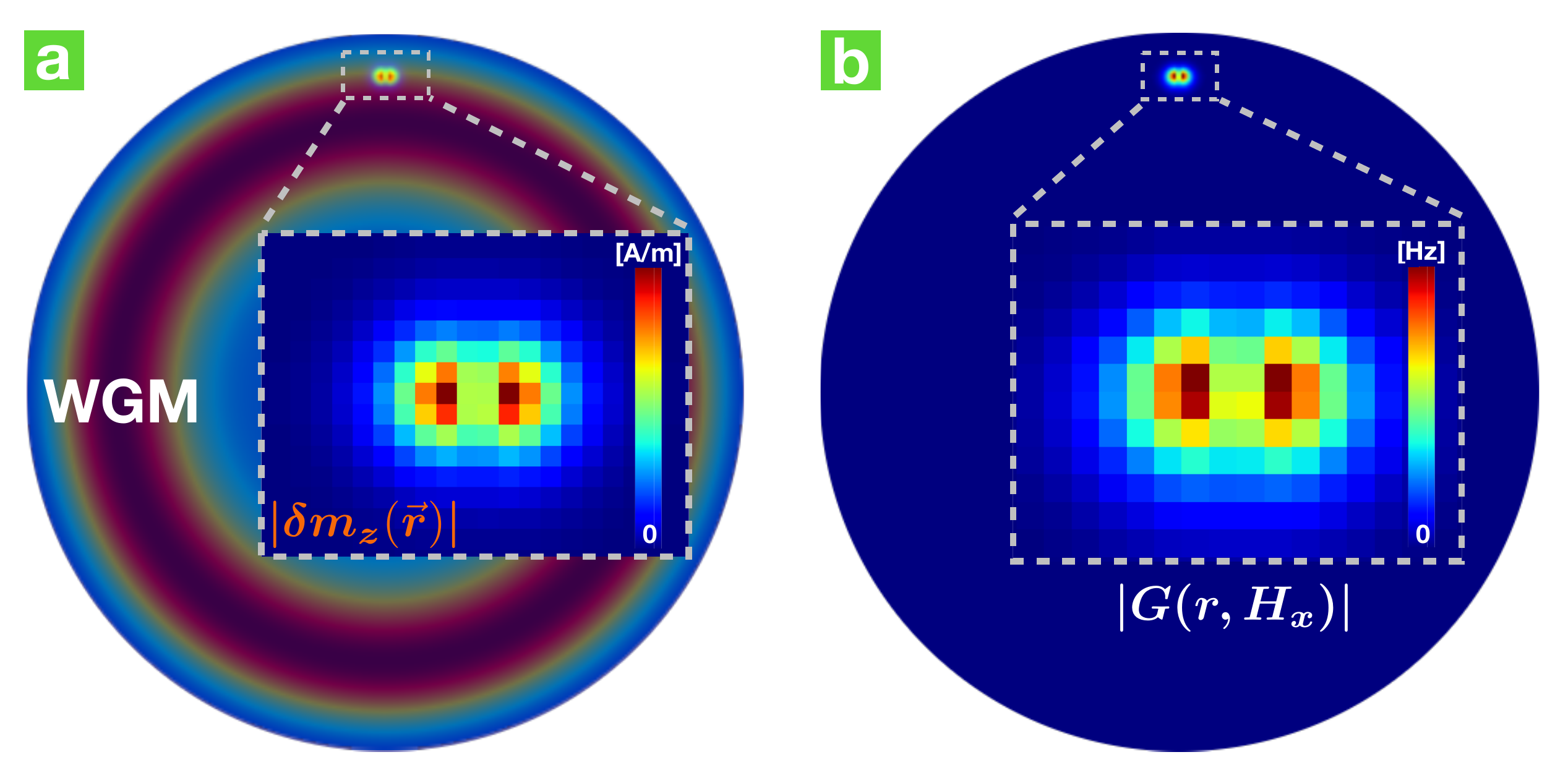}
\par\end{centering}
\caption{Gyrotropic magnon mode and WGM mode and their respective coupling.
(a) WGM of a micromagnetic disk together with the gyrotropic magnon
mode of a displaced magnetic vortex core. The vortex core has been
statically displaced by an applied magnetic field $\mathbf{H}_{x}$
along the $x$ direction. (b) Spatial dependence of the optomagnonic
coupling, $G(\mathbf{r},\mathbf{H}_{x})$ before integrating over
the volume to obtain the to total coupling, $G=\int_{V}{\rm d}^{3}rG(\mathbf{r},\mathbf{H}_{x})$.
The results were obtained with finite element simulations for the
optics and MuMax3 \cite{vansteenkisteDesignVerificationMuMax32014}
micromagnetic simulations for the magnetics, for a thin YIG disk.
In order to confine the optical WGM, a cylindrical heterostructure
consisting of a thin YIG layer between SiN layers was considered,
see Ref. \cite{grafCavityOptomagnonicsMagnetic2018a} for details.}

\label{Fig6}
\end{figure}

We proceed now to calculate an analytical expression for the coupling
of the gyrotropic mode to an optical WGM in the 2D limit. In order to obtain a finite
coupling, the gyrotropic mode needs to have overlap with the WGM,
which lives near the rim of the disk, see Fig. \eqref{Fig6} (a). The
core of the vortex can be displaced from the center of the magnetic
disk by applying an external in-plane magnetic field, as the spins
will try to align with the field. To a first approximation, the position
of the disk $s$ varies linearly with magnetic field, although this
breaks down as the vortex approaches the rim. We will moreover use
the ``rigid vortex'' approximation, which implies that the vortex
moves but does not deform, so that Eqs. \eqref{eq:m_of_rho} are always
valid as a parametrization of the vortex. It is known that this approximation
also fails close to the rim of the disk \cite{guslienkoMagneticVortexState2008}.
Our model therefore is valid for thin disks and its accuracy will
diminish as the vortex approaches the rim of the disk. 

The WGMs in cylindrical geometry, considering as an approximation
an infinite cylinder along $z$, so that the results will be independent
of $z$ due to translation invariance and effectively two-dimensional,
are given by the solution to the Helmholtz equation 
\begin{equation}
\left(\nabla^{2}+n^{2}k_{0}^{2}\right)\psi=0,\label{eq:Helmholtz vortex}
\end{equation}
where $\psi=E_{z}$ or $\psi=B_{z}$ respectively for the TM ($\mathbf{B}\perp\mathbf{e}_{z}$)
and the TE ($\mathbf{E}\perp\mathbf{e}_{z}$) mode, and $n$ is the index
of refraction of the confining dielectric (e.g. YIG). In cylindrical coordinates
$(r,\theta)$ with origin at the center of the disk, the solutions
for $r<R$ ($R$ the radius of the disk) are of the form ($k=nk_{0}$)
\begin{equation}
\psi(r,\theta)=A_{m}J_{m}(kr)e^{i(\pm m\theta)}\,,\label{eq:sol Helm cyl}
\end{equation}
with $J_{m}$ the Bessel function of the first kind together with
the boundary condition 
\begin{equation}
K\partial_{r}J{}_{m}(nkR)/J_{m}(nkR)=\partial_{r}H^{(1)}{}_{m}(kR)/H_{m}^{(1)}(kR)\label{eq:k_mp}
\end{equation}
with $H_{m}^{(1)}$ the Hankel function of the first kind and $K=n$
for a TM and $K=1/n$ for a TE mode. One obtains 
\begin{align}
\mathbf{E}_{mp}^{{\rm TM}}&=\psi_{mp}(r,\theta)\mathbf{e}_{z}\,,\label{eq:TM mode}\\
\mathbf{E}_{mp}^{{\rm TE}}&=\frac{i}{\varepsilon\tilde{\omega}_{mp}}\left(\frac{1}{r}\partial_{\theta}\psi_{mp}\mathbf{e}_{r}-\partial_{r}\psi_{mp}\mathbf{e}_{\theta}\right)\,,\label{eq:TE mode}
\end{align}
where the tilde indicates that the solutions are complex: $\tilde{\omega}=\frac{c}{n}\,\tilde{k}=\omega+\frac{i\kappa}{2}$,
and the subscripts $mp$ indicates that the expressions are evaluated
for a particular solution $\tilde{k}=\tilde{k}_{mp}=k_{mp}-ik_{mp}''$
of Eq. \eqref{eq:k_mp}. In the following we consider only WGMs solutions,
which correspond to $p=1$ (one node in the radial direction), hence
we will omit this index. Well defined WGMs are solutions with small
imaginary part $k_{m1}''$, since this is related to the leaking of
the optical mode out of the cavity. The imaginary part gives the decay
rate of the mode due to coupling to external unbounded optical modes,
and will enter in the total decay rate of the mode. The normalization
of the WGM can found by imposing Eq. \eqref{eq:E_norm}.

\begin{figure}
\begin{centering}
\includegraphics[width=0.8\textwidth]{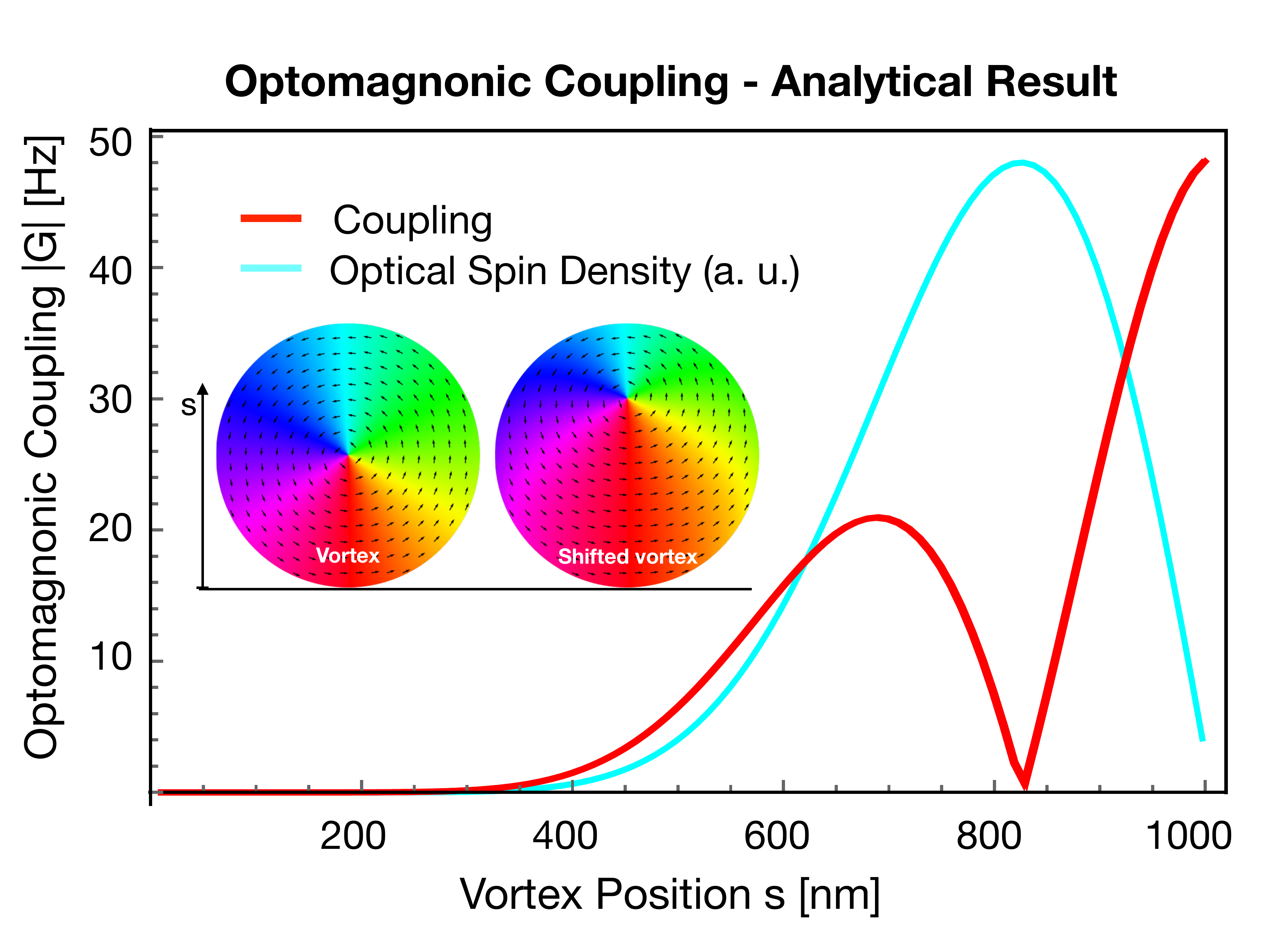}
\par\end{centering}
\caption{Optomagnonic coupling (red) as a function of the vortex position according
to Eq. \eqref{eq:G_vortex}. Also shown is the (normalized) magnitude
of the optical spin density (see Eq. \eqref{eq:OSD}) at the position
of the vortex. From the plot it is evident that the coupling is proportional
to the gradient of the optical spin density. The inset shows the magnetic
vortex at zero field and at an arbitrary finite magnetic field along
$x$. Radius of the YIG thin disk $R=1\mu{\rm m}$), WGM with $m=6$,
$\omega_{{\rm opt}}/2\pi\approx200{\rm THz}$. Adapted from Ref. \cite{grafCavityOptomagnonicsMagnetic2018a}.}
 \label{Fig7}
\end{figure}

We consider for simplicity the optomagnonic coupling to only one WGM.
One can easily show that the only possibility for finite coupling
in the 2D limit is to couple to the TE mode \cite{grafCavityOptomagnonicsMagnetic2018a}.
It is straightforward to obtain
\begin{equation}
G_{m}\!=\!-i\frac{\theta_{{\rm F}}\lambda_{n}}{4\pi}\frac{\varepsilon_{0}\varepsilon}{2}h\!\int_{0}^{b}\!\!\rho d\rho\int_{0}^{2\pi}\!\!d\varphi m_{z}(\rho,\varphi)\left(\mathbf{E}_{m}^{{\rm TE}*}\times\mathbf{E}_{m}^{{\rm TE}}\right)\cdot\mathbf{e}_{z}\label{eq:G ana}
\end{equation}
where $(\rho,\varphi)$ are polar coordinates in the system with origin
at the center of the vortex (note that the center of the vortex can
be displaced by an external magnetic field, see Fig. \eqref{Fig6}).
From Eq.\eqref{eq:TE mode} we obtain 
\begin{equation}
G_{\pm m}\!=\!\pm\frac{\theta_{{\rm F}}\lambda_{n}}{2\pi}\frac{\hbar\omega_{m}}{2\pi\mathcal{N}_{J}}r_{c}k_{m}m\int_{0}^{1}d\rho\int_{0}^{2\pi}d\varphi e^{i\varphi}\frac{\rho^{2}}{\left(\rho^{2}+1\right)^{2}}\frac{\partial_{\tilde{r}}|J_{m}(\tilde{r})|^{2}}{|\left(s/b\right)\mathbf{e}_{y}+\rho\mathbf{e}_{\rho}|}\,,\label{eq:G_vortex}
\end{equation}
with $\tilde{r}=k_{m}r$, $\mathbf{r}=s\mathbf{e}_{y}+\rho\mathbf{e}_{\rho}$,
and $\omega_{m}=\frac{c}{n}k_{m}$. $r_{c}$ and $\mathcal{N}_{J}$
are the corresponding normalization factors for the magnon and optical
modes and depend on the geometrical parameters of the system. An example
of the spatial structure of the coupling \eqref{eq:G_vortex} is shown
in Fig. \eqref{Fig6} (b), where the results were obtained via micromagnetics \cite{vansteenkisteDesignVerificationMuMax32014}
and finite element simulations for the optics. One can show that the
first order contribution to Eq. \eqref{eq:G_vortex} is proportional
to the gradient of the optical spin density with respect to the vortex
position. This can be also seen in the results of Fig. \eqref{Fig7}.

The 2D approximation works quite well for a thin YIG disk, as we have
shown in Ref. \cite{grafCavityOptomagnonicsMagnetic2018a}, where
we compared the analytical results with results obtained by combining
micromagnetic and finite element simulations. Whereas the thin disk
is perfectly fine to host the magnon modes, it is a bad optical cavity
since it is not able to confine the light. As a rule of thumb, the
thickness of the dielectric has to be of the order of at least half
of the wavelength of the light in the material in order to confine
it effectively. For this reason, we proposed a ``sandwich'' heterostructure,
where the YIG thin disk is sandwiched between thicker layers of SiN
which serve to confine the light. The index of refraction of the transparent
dielectric SiN is similar to that of YIG in the optical range, so
it is a good choice of material. This however has the effect of decreasing
the magnon-to-optical volume ratio, which is detrimental for the optomagnonic
coupling as discussed previously. A solution to enhance the coupling
was proposed in Ref. \cite{grafCavityOptomagnonicsMagnetic2018a},
where instead of a thin disk a thicker YIG structure was considered.
In this case however, the 2D approximation breaks down and the results
for the coupling necessarily must be obtained numerically. Promising
high values for the coupling and cooperativity were obtained, indicating
the value of studying and designing optomagnonic systems beyond the
homogeneous, Kittel mode case. 

\chapter{A quantum protocol: all-optical magnon heralding}

To finish this chapter, we study a possible quantum protocol in a
cavity optomagnonic system, as proposed in Ref. \cite{bittencourtMagnonHeraldingCavity2019}.
In the protocol, a one-magnon Fock state is created in the magnetic
material by the detection of a photon. This is referred to as \emph{heralding},
since the detected photon announces the creation of the desired state.
A Fock state, also called a number state since it has a well-defined
occupation number, is a purely nonclassical state (as compared for
example with a coherent state) characterized by a negative Wigner
function. A magnon Fock state is, therefore, a macroscopic collective
(involving millions of spins) nonclassical state of the magnetic
system, and its realization can be the first step towards the generation, manipulation,
and transfer of quantum states in optomagnonic systems \cite{lachance-quirionResolvingQuantaCollective2017,lachance-quirionEntanglementbasedSingleshotDetection2019}.
Our protocol proposal includes generating the magnon Fock state optically
and reading the state, also optically, at some time later. Due to
the optomagnonic interaction, photons and magnons are entangled
during the evolution, and detecting a photon projects the magnon sate.
The successful generation of the nonclassical state is tested by
measuring the two-photon correlations of a ``reading'' laser. We consider
a system with two optical modes and one relevant magnon mode, in line
with current experimental setups involving optical whispering gallery
modes in YIG spheres \cite{haighTripleResonantBrillouinLight2016,zhangOptomagnonicWhisperingGallery2016,osadaCavityOptomagnonicsSpinOrbit2016}.

\section{Hamiltonian and Langevin Equations of Motion}

In this section, we analyze the quantum Langevin equations of motion
of the cavity optomagnonic system in the spin-wave regime for a system
with two non-degenerate optical modes $\hat{a}_{1}$ and $\hat{a}_{2}$
interacting with one magnon mode $\hat{m}$. We will find the analytical
solutions for the evolution of the quantum fields by linearizing also
in the optical fields, as in Eq. \eqref{eq:linSHMO}. Our analysis
is valid for any magnetization texture and magnon mode (including
the homogeneous case), but it is restricted to small oscillations
of the spins.

In this case the Hamiltonian in Eq. \eqref{eq:H_tot_SW} reduces to
\begin{eqnarray}
\hat{H} & = & -\hbar\Delta_{1}\hat{a}_{1}^{\dagger}\hat{a}_{1}-\hbar\Delta_{2}\hat{a}_{2}^{\dagger}\hat{a}_{2}\label{eq:Ham2Mfull}\\
 & + & {\displaystyle \hbar\Omega\hat{m}^{\dagger}\hat{m}+\hbar}\left(G_{12}\hat{a}_{1}^{\dagger}\hat{a}_{2}+G_{21}\hat{a}_{2}^{\dagger}\hat{a}_{1}\right)\hat{m}^{\dagger}+h.c.\nonumber \\
 & + & i\sum_{j}\epsilon_{j}(\hat{a}_{j}^{\dagger}-\hat{a}_{j})\,,
\end{eqnarray}
where we consider that modes $j=1,\,2$ are driven at frequency $\omega_{{\rm L}}$
and amplitude $\epsilon_{j}$ given by Eq. \eqref{eq:powerEps}, and
$\Delta_{j}=\omega_{L}-\omega_{j}$ is the respective detuning. The
particularity of the cavity optomagnonic system involving spherical
WGMs is the asymmetry between the scattering rates $G_{12}$ and $G_{21}$,
due to energy and angular momentum conservation rules \cite{osadaCavityOptomagnonicsSpinOrbit2016}.
In particular, one of the two processes is highly suppressed, which
we reflect by setting
\begin{equation}
G_{21}=0=G_{21}^{*}\label{eq:G12 0}
\end{equation}
in Eq. \eqref{eq:Ham2Mfull}. Accordingly, we consider two optical
modes satisfying approximately the resonance condition $\omega_{2}-\omega_{1}\approx\Omega$.
Applying Eq. \eqref{eq:heis EOM} to each of the field operators,
and including dissipative and noise terms both for photons and magnons
following Eq. \eqref{eq:LEOMinout}, we obtain 

\begin{eqnarray}
\frac{d\hat{a}_{1}}{dt} & = & i\Delta_{1}\hat{a}_{1}-iG_{12}{\displaystyle \hat{a}_{2}}\hat{m}^{\dagger}-\frac{\kappa}{2}\hat{a}_{1}+\sqrt{\kappa}\,\hat{a}_{{\rm 1,in}}(t)+\epsilon_{1}\,,\label{eq:a1a2m}\\
\frac{d\hat{a}_{2}}{dt} & = & i\Delta_{2}\hat{a}_{2}-iG_{12}^{*}{\displaystyle \hat{a}_{1}}\hat{m}-\frac{\kappa}{2}\hat{a}_{2}+\sqrt{\kappa}\,\hat{a}_{{\rm 2,in}}(t)+\epsilon_{2}\,,\nonumber \\
\frac{d\hat{m}}{dt} & = & -i\Omega\hat{m}-iG_{12}\hat{a}_{1}^{\dagger}\hat{a}_{2}-\frac{\Gamma}{2}\hat{m}++\sqrt{\Gamma}\hat{m}_{{\rm in}}(t)\,,\nonumber 
\end{eqnarray}
where we have assumed for simplicity that both photon modes are subject
to the same decay rate $\kappa$. Whereas the photon bath can be considered
to be at zero temperature due to the high frequency of the optical
photons,
\begin{align}
\langle\hat{a}_{i,{\rm in}}(t)\hat{a}_{j,{\rm in}}^{\dagger}(t^{\prime})\rangle & =\delta_{ij}\delta(t-t^{\prime})\,,\label{eq:Noise12}\\
\langle\hat{a}_{i,{\rm in}}^{\dagger}(t)\hat{a}_{j,{\rm in}}(t)\rangle & =0\,,\nonumber 
\end{align}
the magnons have usually ${\rm GHz}$ frequencies and the temperature
of the thermal bath cannot be ignored, unless the system is cooled
to mK temperatures. The general expressions for the magnon
correlators are

\begin{eqnarray}
\langle\hat{m}_{{\rm in}}(t)\hat{m}_{{\rm in}}^{\dagger}(t^{\prime})\rangle & = & (n_{{\rm tm}}+1)\delta(t-t^{\prime}),\label{eq:NoiseMagnon}\\
\langle\hat{m}_{{\rm in}}^{\dagger}(t)\hat{m}_{{\rm in}}(t^{\prime})\rangle & = & n_{{\rm {\rm tm}}}\delta(t-t^{\prime})\,,
\end{eqnarray}
where the subscript ${\rm tm}$ indicates that $n_{{\rm {\rm tm}}}$
is the mean number of thermal magnons given by the Bose-Einstein distribution
\begin{equation}
n_{{\rm {\rm tm}}}(T)=\frac{1}{\exp\left(\hbar\,\Omega/k_{B}T\right)-1}\,,\label{eq:n tm}
\end{equation}
where $T$ is the temperature of the magnon bath and $k_{B}$ the
Boltzmann constant, and $\Omega$ is the magnon mode frequency.
\[
\]

The steady state values $\langle\hat{a}_{i}\rangle=\alpha_{i}$ and
$\langle\hat{m}\rangle=\beta$ are found by setting Eqs. \eqref{eq:a1a2m}
to zero and ignoring the noise fluctuations. The expectation values
of the operators are understood to be taken with respect to a coherent
state. In the side-band resolved limit ( $\Omega\gg\kappa,\gamma$
) it is straightforward to see that if only one mode $j$ (=1 or 2)
is driven, $\beta=0$ and the steady state circulating number of photons
in the cavity is given by $|\alpha_{j}|^{2}$ with 

\begin{equation}
\alpha_{j}=-\frac{\epsilon_{j}}{i\Delta_{j}-\kappa_{j}/2}.\label{eq:alpha j}
\end{equation}

By considering the fluctuations around the steady state $\hat{a}_{i}\rightarrow\alpha_{i}+\hat{a}_{i}$,
$\hat{m}\rightarrow\hat{m}$ (where for simplicity of notation we
denote now \emph{the fluctuations} by $\hat{a}_{i}$ and $\hat{m}$)
one obtains the Hamiltonian valid in the linear regime, as in Eq.
\eqref{eq:linSAHMO}. In the interaction picture the resulting Hamiltonian
reads
\begin{align}
\hat{H}_{{\rm IP}} & \approx\hbar\alpha_{1}^{*}G_{12}\hat{a}_{2}\hat{m}^{\dagger}e^{i(\Delta_{2}+\Omega)t}+\hbar\alpha_{2}G_{12}\hat{a}_{1}^{\dagger}\hat{m}^{\dagger}e^{-i(\Delta_{1}-\Omega)t}+h.c.\,.\label{eq:HIPfull}
\end{align}

\section{Write and read protocol}

From Eq. \eqref{eq:HIPfull} one can immediately observe that by pumping
the optical mode 2 at resonance ($\omega_{{\rm L}}=\omega_{2}\approx\omega_{1}+\Omega$)
while mode 1 is not driven ($\alpha_{1}=0$), the condition $\Delta_{1}=\Omega$
is satisfied and one obtains
\begin{equation}
\hat{H}_{{\rm {\rm W}}}\approx\hbar\left[\alpha_{2}^{*}G_{12}^{*}\hat{a}_{1}\hat{m}+\alpha_{2}G_{12}\hat{a}_{1}^{\dagger}\hat{m}^{\dagger}\right]\,,\label{eq:Hwrite}
\end{equation}
in which a magnon and a photon in mode 1 are either created or annihilated
in pairs. We denote this effective Hamiltonian as $\hat{H}_{{\rm {\rm W}}}$
since it corresponds to the writing Hamiltonian is our protocol. Starting
from the vacuum of magnons and photons, the evolution under this Hamiltonian
creates entangled pairs of magnons and mode-1 photons. Detecting a
photon in mode 1 collapses the entangled state and determines the
magnon state, with a certain probability of collapsing into a one-magnon
Fock state\textbf{ }\cite{bittencourtMagnonHeraldingCavity2019}.
The Fock state created in this form is denominated \emph{heralded}. 

\begin{figure}
\begin{centering}
\includegraphics[width=0.8\textwidth]{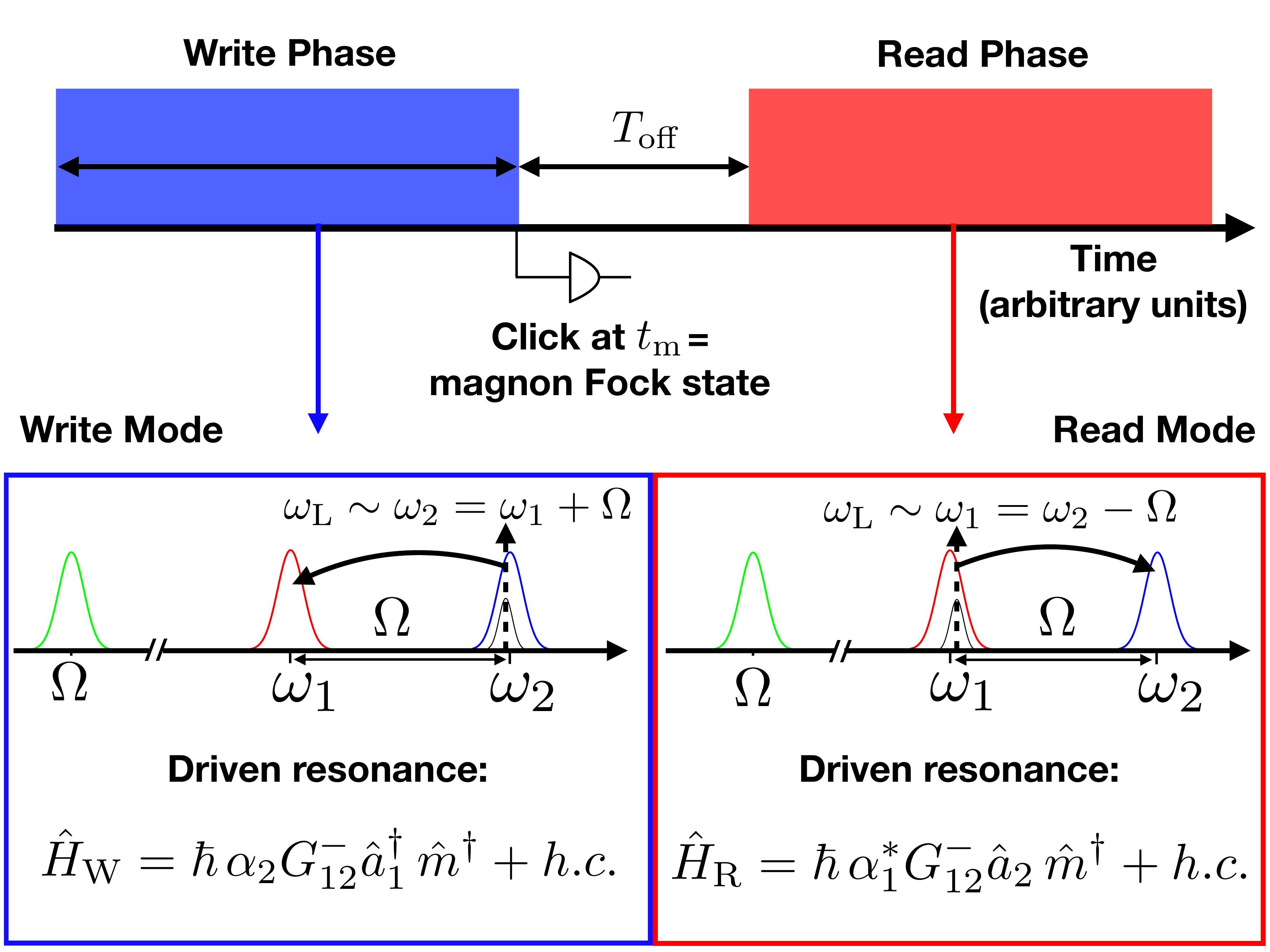}
\par\end{centering}
\caption{Write and read protocol for heralding a magnon Fock state. The detection
of a photon at time $t_{{\rm m}}$ is a probabilistic process. Figure
taken from Ref. \cite{bittencourtMagnonHeraldingCavity2019}.}

\label{Fig8}
\end{figure}

The successful heralding of a magnon Fock state can be corroborated
by a reading protocol, as long this is done within the lifetime of
the state, that is , the lifetime of the magnon. The reading Hamiltonian
is obtained from Eq. \eqref{eq:HIPfull} by this time driving the
optical mode 1 at resonance ($\omega_{{\rm L}}=\omega_{1}\approx\omega_{2}-\Omega$)
and not driving mode 2 ($\alpha_{2}=0$). One obtains
\begin{equation}
\hat{H}_{{\rm {\rm R}}}\approx\hbar\left[\alpha_{1}^{*}G_{12}\hat{a}_{2}\hat{m}^{\dagger}+\alpha_{1}G_{12}^{*}\hat{a}_{2}^{\dagger}\hat{m}\right]\,.\label{eq:Hread}
\end{equation}
In the strong coupling regime, such that the effective coupling strength
$G_{1,12}^{{\rm eff}}=\vert\alpha_{1}G_{12}\vert$ is larger than
the decay channels $\vert\alpha_{1}G_{12}\vert>\kappa,\,\Gamma$,
the magnon mode and the photon mode hybridize, giving rise to eigenmodes
which are part magnon, part photon, in analogy with the cavity magnon
polariton discussed previously in the MW regime. Moreover, if $\vert\alpha_{1}G_{12}\vert>\kappa,\,n_{{\rm tm}}\Gamma$
(denominated the coherent coupling regime) the interaction is quantum-coherent,
allowing to transfer the magnon state coherently to the photons. Therefore
measuring the mode-2 photons give information on the magnon state,
that is, we can ``read'' the state. The write-and-read protocol is
described schematically in Fig. \eqref{Fig8}.

As we pointed out in the introduction, the strong coupling regime
in optomagnonics is challenging to attain and experiments have not
yet reached this point. Nevertheless, photons in mode 2 can be used
to probe the heralded state even in the weak coupling regime. This
can be done by measuring the two-photon correlation function \cite{wallsQuantumOptics2008}
\begin{equation}
g_{{\rm Read}}^{(2)}(t,t+\tau)=\frac{\langle\hat{a}_{2}^{\dagger}(t)\hat{a}_{2}^{\dagger}(t+\tau)\hat{a}_{2}(t+\tau)\hat{a}_{2}(t)\rangle_{{\rm }}}{\langle\hat{a}_{2}^{\dagger}(t)\hat{a}_{2}(t)\rangle_{{\rm }}\langle\hat{a}_{2}^{\dagger}(t+\tau)\hat{a}_{2}(t+\tau)\rangle_{{\rm }}}\,,\label{eq:g2def}
\end{equation}
which can be done interferometrically. In order to use this quantity
as a \emph{heralding witness}, the expectation values have to be taken
after the measurement of a ``write'' photon (a mode-1 photon). Given
the form of the read Hamiltonian from Eq. \eqref{eq:Hread}, which
``swaps'' magnons with photons, measuring $g_{{\rm Read}}^{(2)}$
for the photons is equivalent to measure the corresponding correlation
function for the magnons. If the state is non-classical, $g_{{\rm Read}}^{(2)}(t,t)<1$,
given an indication that the heralded magnon state is a Fock state.
The condition $g_{{\rm Read}}^{(2)}(t,t)<1$ is denominated antibunching,
since there is a reduced probability of two photons being detected
simultaneously, and it is an example of a quantum state violating
a classical inequality ($g_{{\rm Read}}^{(2)}(t,t)>1$ necessarily
for classical states). The reliability of $g_{{\rm Read}}^{(2)}$
as a true heralding witness depends however on the temperature of the
magnon bath, worsening as the temperature and consequently the number
of thermal magnons increases \textbf{\cite{bittencourtMagnonHeraldingCavity2019}}.

\section{Solution of the linear quantum Langevin equations}

The probability of heralding a magnon, as well as the correlation
function $g_{{\rm Read}}^{(2)}$, can be calculated by the linear
quantum Langevin equations dictated by the Hamiltonians of Eq. \eqref{eq:Hwrite}
and \eqref{eq:Hread} with the steady state given by Eqs. \eqref{eq:a1a2m}
and using the noise correlators of Eqs. \eqref{eq:Noise12} and \eqref{eq:NoiseMagnon}.
The linear Langevin equations can be written in compact form as
\begin{equation}
\frac{d\hat{\bm{A}}}{dt}=M^{{\rm X}}\cdot\hat{\bm{A}}(t)+\bm{\hat{N}}(t),\label{eq:EOM A}
\end{equation}
where
\begin{eqnarray}
\hat{\bm{A}}=\left(\begin{array}{c}
\hat{a}_{1}^{\dagger}\\
\hat{m}\\
\hat{a}_{2}
\end{array}\right), &  & \hat{\bm{N}}=\left(\begin{array}{c}
\sqrt{\kappa}(\hat{a}_{1}^{in})^{\dagger}\\
\sqrt{\gamma}\hat{m}^{in}\\
\sqrt{\kappa}\hat{a}_{2}^{in}
\end{array}\right),\label{eq:DefinitionVectors}
\end{eqnarray}
${\rm X}={\rm W},$ ${\rm R}$ indicates the write or read phases
of the protocol, and $M^{{\rm X}}$ is a matrix given by the Heisenberg
equation of motion deriving from Hamiltonians \eqref{eq:Hwrite} or
\eqref{eq:Hread} . The solution can be written formally as 
\begin{equation}
\hat{\bm{A}}(t)=U^{{\rm X}}(t)\cdot\hat{\bm{A}}(0)+{\displaystyle \int}_{0}^{t}d\tau U^{{\rm X}}(t-\tau)\cdot\hat{\bm{N}}(\tau),\label{eq:SolutionsLangevin}
\end{equation}
where $U^{{\rm W}}(t)$ and $U^{{\rm R}}(t)$ are the respective evolution
matrices. These can be found analytically by going to a diagonal basis
such that 
\begin{equation}
\frac{d\hat{A}_{i}^{\prime}}{dt}=\lambda_{i}^{{\rm P}}\hat{A}_{i}^{\prime}(t)+\hat{N}_{i}^{\prime}(t),\label{eq:EOM A diagonal}
\end{equation}
which can be easily integrated and transformed back to find $U^{{\rm W}}(t)$
and $U^{{\rm R}}(t)$. The expressions for $U^{{\rm W}}(t)$ and $U^{{\rm R}}(t)$
are lengthy and can be found in Ref. \cite{bittencourtMagnonHeraldingCavity2019}.

\section{Probability of heralding a magnon}

\begin{figure}
\begin{centering}
\includegraphics[width=1\textwidth]{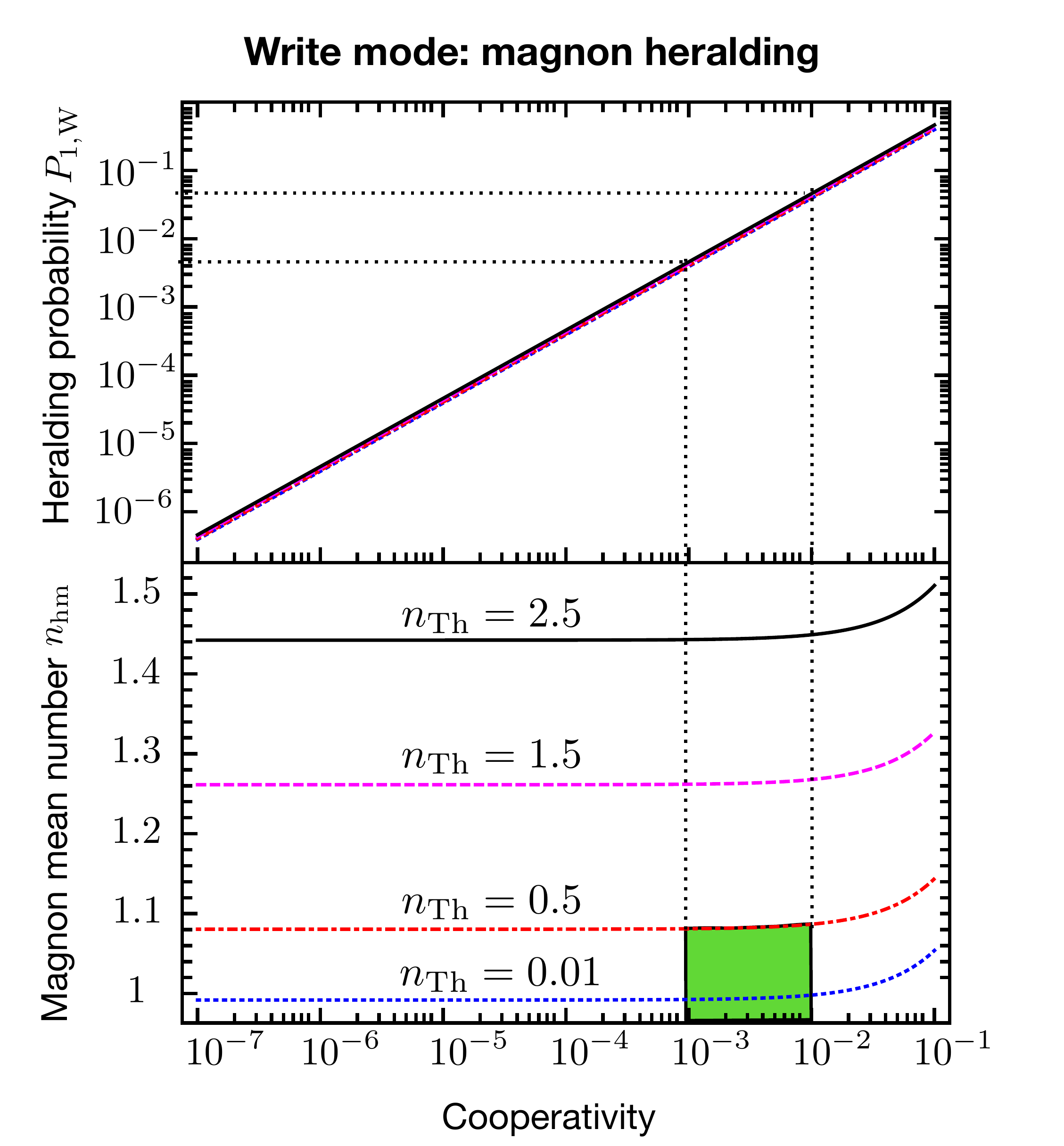}
\par\end{centering}
\caption{Probability of heralding a magnon and mean number of heralded magnons
as a function of the write cooperativity, $\mathcal{C}=4G_{1,12}^{{\rm eff}}/\kappa\Gamma$.
The shaded area indicates values of the cooperativity which give an
appreciable heralding probability while keeping the number of heralded
magnons near one. The number of thermal magnons $n_{{\rm tm}}$ is
dictated by the temperature of the magnon bath, larger temperatures
are, as expected, detrimental for the heralding protocol. Figure taken
from Ref. \cite{bittencourtMagnonHeraldingCavity2019}.}

\label{Fig9}
\end{figure}

The probability of heralding a magnon is given by the probability
of detecting a photon in mode 1 during the write phase. This is given
by (see e.g. Ref. \cite{wallsQuantumOptics2008})
\[
P_{1,{\rm W}}(t)=\langle:\hat{a}_{1}^{\dagger}\hat{a}_{1}\exp(-\hat{a}_{1}^{\dagger}\hat{a}_{1}):\rangle\sim\langle\hat{a}_{1}^{\dagger}\hat{a}_{1}\rangle-\langle\hat{a}_{1}^{\dagger}\hat{a}_{1}^{\dagger}\hat{a}_{1}\hat{a}_{1}\rangle\,,
\]
where the approximation is valid as long as $\langle\hat{a}_{1}^{\dagger}\hat{a}_{1}^{\dagger}\hat{a}_{1}^{\dagger}\hat{a}_{1}\hat{a}_{1}\hat{a}_{1}\rangle\ll\langle\hat{a}_{1}^{\dagger}\hat{a}_{1}^{\dagger}\hat{a}_{1}\hat{a}_{1}\rangle$.
This probability can be computed by the method outlined previously,
on the basis of the solution for the equations of motion \eqref{eq:EOM A}
for the write-Hamiltonian. One finds that the heralding probability
grows linearly with the square of the effective optomagnonic coupling
during the write phase. A very large coupling is however detrimental,
since it will tend to generate a larger number of magnons within the
same period, compromising the Fock state. Therefore an optimal effective
coupling strength $G_{1,12}^{{\rm eff}}$ must be found, see Fig.
\eqref{Fig9}. The mean number of magnons $n_{{\rm hm}}=\langle\hat{m}^{\dagger}\hat{m}\rangle$
after the heralding event depends crucially on the temperature of
the magnon bath, as shown in Fig. \eqref{Fig9}.

\section{Magnon cooling}

As we have seen in the heralding example, the presence of thermal
magnons is highly detrimental for quantum protocols. Due to their
larger frequencies (usually in the ${\rm GHz}$ range in optomagnonic
setups), magnons are generally more amenable to direct cooling than,
for example, phonons in optomechanical systems. However light can
be used for active cooling in cases in which direct cooling is not
sufficient or inconvenient from and experimental point of view. Magnon
cooling in optomagnonic systems has been studied in detail in Ref.
\cite{sharmaOpticalCoolingMagnons2018}. 

The formalism described to solve the quantum Langevin equations of
motion can be used to study the cooling protocol exactly, including
all decay channels and quantum fluctuations. In particular, our ``read''
Hamiltonian can be used for magnon cooling, since it effectively annihilates
magnons when mode 1 is driven. If we consider an initial thermal state
state with a mean number of magnons $n_{{\rm th}}$, the initial density
matrix of the optomagnonic system is given by 
\begin{eqnarray}
\rho(0) & = & \vert0\rangle\langle0\vert_{1}\otimes\vert0\rangle\langle0\vert_{2}\otimes\rho_{{\rm th,m}},\label{eq:rho}
\end{eqnarray}
where 
\begin{equation}
\rho_{{\rm th,\,m}}=\frac{1}{1+n_{{\rm th}}}\sum_{n\ge0}\left[\frac{n_{{\rm th}}}{1+n_{{\rm th}}}\right]^{n}\vert n\rangle\langle n\vert,\label{eq:rho th}
\end{equation}
indicates that the initial population of magnons os a thermal state.
The temporal evolution of the mean number of magnons is given by 
\begin{eqnarray}
\langle\hat{m}^{\dagger}\hat{m}\rangle(t) & =\sum_{i,j} & \left(U_{i2}^{{\rm R}}(t)\right)^{*}U_{2j}^{{\rm R}}(t)\langle\hat{A}_{i}^{\dagger}(0)\hat{A}_{j}(0)\rangle\label{eq:n_m(t)}\\
 & + & \sum_{i,j}{\displaystyle \int}_{0}^{\,t}d\tau_{1}d\tau_{2}\left(U_{i2}^{{\rm R}}(t-\tau_{1})\right)^{*}U_{2j}^{{\rm R}}(t-\tau_{2})\langle\hat{N}_{i}^{\dagger}(\tau_{1})\hat{N}_{j}(\tau_{2})\rangle\,,\nonumber 
\end{eqnarray}
where $i,j$ indicate the components of $\hat{\mathbf{A}}$, $\hat{\mathbf{N}}$,
and $U^{{\rm R}}$ as defined in Eqs. \eqref{eq:DefinitionVectors}
and \eqref{eq:SolutionsLangevin}, and the expectation values are
taken over the initial state determined by $\rho(0)$. Imposing the
noise correlators of Eqs. \eqref{eq:Noise12} and \eqref{eq:NoiseMagnon}
one obtains
\begin{eqnarray}
\langle\hat{m}^{\dagger}\hat{m}\rangle(t) & = & \vert U_{22}^{{\rm R}}(t)\vert^{2}n_{{\rm th}}+\vert U_{12}^{{\rm R}}(t)\vert^{2}\label{eq:n m t 2}\\
 & + & \int_{0}^{t}d\tau\left[\vert U_{12}^{{\rm R}}(t-\tau)\vert^{2}+\vert U_{22}^{{\rm R}}(t-\tau)\vert^{2}n_{{\rm th}}\right].\nonumber 
\end{eqnarray}
The steady state value $n_{0}=\langle\hat{m}^{\dagger}\hat{m}\rangle(t\rightarrow\infty)$
can be shown to be
\[
n_{0}=\frac{\Gamma n_{{\rm th}}}{(\kappa+\Gamma)}\left(1+\frac{\kappa}{\Gamma(1+C_{{\rm R}})}\right),
\]
where $C_{{\rm R}}=4|\alpha_{1}^{*}G_{1{\rm 2}}|^{2}/\kappa\Gamma$
is the read-phase cooperativity. This is a thermal state with $n_{0}<n_{{\rm th}}$,
and therefore it is cooled. Whereas active cooling is necessary for
most implementations of optomechanical systems, given that phonon
frequencies are usually low, for optomagnonic systems this can be
circumvented by cooling the system by usual dilution fridge refrigeration
techniques.

\chapter{Outlook}

Cavity optomagnonic systems are at the interface between condensed
matter and quantum optics, and present new opportunities to study
and control the interaction between light and magnetic systems, in
particular at the single quanta level. Magnons are robust elementary
excitations, highly tunable and can couple well to several other degrees
of freedom beyond photons, such as phonons and electrons. The incorporation
of magnetically ordered systems into hybrid platforms for quantum
information is therefore very promising. From a fundamental point
of view, cavity optomagnonic systems are very rich. Topics such as
the interaction of structured light with magnetic textures and topological
defects, the nonlinear dynamics of the coupled system, or collective
quantum effects in cavity optomagnonics taking into account the strongly
correlated nature of the magnetically ordered systems, are still largely
unexplored. There are exciting times ahead for theorists and experimentalists
alike.

\bibliographystyle{unsrt}
\bibliography{BookOptomagnonicStructures}

\begin{thebibliography}{10}

\bibitem{macfarlaneQuantumTechnologySecond2003}
A.~G.~J. MacFarlane, Jonathan~P. Dowling, and Gerard~J. Milburn.
\newblock Quantum technology: The second quantum revolution.
\newblock {\em Philosophical Transactions of the Royal Society of London.
  Series A: Mathematical, Physical and Engineering Sciences},
  361(1809):1655--1674, August 2003.

\bibitem{aruteQuantumSupremacyUsing2019}
Frank Arute, Kunal Arya, Ryan Babbush, Dave Bacon, Joseph~C. Bardin, Rami
  Barends, Rupak Biswas, Sergio Boixo, Fernando G. S.~L. Brandao, David~A.
  Buell, Brian Burkett, Yu~Chen, Zijun Chen, Ben Chiaro, Roberto Collins,
  William Courtney, Andrew Dunsworth, Edward Farhi, Brooks Foxen, Austin
  Fowler, Craig Gidney, Marissa Giustina, Rob Graff, Keith Guerin, Steve
  Habegger, Matthew~P. Harrigan, Michael~J. Hartmann, Alan Ho, Markus Hoffmann,
  Trent Huang, Travis~S. Humble, Sergei~V. Isakov, Evan Jeffrey, Zhang Jiang,
  Dvir Kafri, Kostyantyn Kechedzhi, Julian Kelly, Paul~V. Klimov, Sergey Knysh,
  Alexander Korotkov, Fedor Kostritsa, David Landhuis, Mike Lindmark, Erik
  Lucero, Dmitry Lyakh, Salvatore Mandr{\`a}, Jarrod~R. McClean, Matthew
  McEwen, Anthony Megrant, Xiao Mi, Kristel Michielsen, Masoud Mohseni, Josh
  Mutus, Ofer Naaman, Matthew Neeley, Charles Neill, Murphy~Yuezhen Niu, Eric
  Ostby, Andre Petukhov, John~C. Platt, Chris Quintana, Eleanor~G. Rieffel,
  Pedram Roushan, Nicholas~C. Rubin, Daniel Sank, Kevin~J. Satzinger, Vadim
  Smelyanskiy, Kevin~J. Sung, Matthew~D. Trevithick, Amit Vainsencher, Benjamin
  Villalonga, Theodore White, Z.~Jamie Yao, Ping Yeh, Adam Zalcman, Hartmut
  Neven, and John~M. Martinis.
\newblock Quantum supremacy using a programmable superconducting processor.
\newblock {\em Nature}, 574(7779):505--510, October 2019.

\bibitem{kimbleQuantumInternet2008}
H.~J. Kimble.
\newblock The quantum internet.
\newblock {\em Nature}, 453:1023--1030, June 2008.

\bibitem{oconnellQuantumGroundState2010a}
A.~D. O'Connell, M.~Hofheinz, M.~Ansmann, Radoslaw~C. Bialczak, M.~Lenander,
  Erik Lucero, M.~Neeley, D.~Sank, H.~Wang, M.~Weides, J.~Wenner, John~M.
  Martinis, and A.~N. Cleland.
\newblock Quantum ground state and single-phonon control of a mechanical
  resonator.
\newblock {\em Nature}, 464(7289):697--703, April 2010.

\bibitem{chanLaserCoolingNanomechanical2011a}
Jasper Chan, T.~P.~Mayer Alegre, Amir~H. {Safavi-Naeini}, Jeff~T. Hill, Alex
  Krause, Simon Gr{\"o}blacher, Markus Aspelmeyer, and Oskar Painter.
\newblock Laser cooling of a nanomechanical oscillator into its quantum ground
  state.
\newblock {\em Nature}, 478(7367):89--92, October 2011.

\bibitem{teufelSidebandCoolingMicromechanical2011a}
J.~D. Teufel, T.~Donner, Dale Li, J.~W. Harlow, M.~S. Allman, K.~Cicak, A.~J.
  Sirois, J.~D. Whittaker, K.~W. Lehnert, and R.~W. Simmonds.
\newblock Sideband cooling of micromechanical motion to the quantum ground
  state.
\newblock {\em Nature}, 475(7356):359--363, July 2011.

\bibitem{aspelmeyerCavityOptomechanics2014}
Markus Aspelmeyer, Tobias~J. Kippenberg, and Florian Marquardt.
\newblock Cavity optomechanics.
\newblock {\em Rev. Mod. Phys.}, 86(4):1391--1452, December 2014.

\bibitem{riedingerRemoteQuantumEntanglement2018}
Ralf Riedinger, Andreas Wallucks, Igor Marinkovi{\'c}, Clemens L{\"o}schnauer,
  Markus Aspelmeyer, Sungkun Hong, and Simon Gr{\"o}blacher.
\newblock Remote quantum entanglement between two micromechanical oscillators.
\newblock {\em Nature}, 556(7702):473--477, April 2018.

\bibitem{kurizkiQuantumTechnologiesHybrid2015}
Gershon Kurizki, Patrice Bertet, Yuimaru Kubo, Klaus M{\o}lmer, David
  Petrosyan, Peter Rabl, and J{\"o}rg Schmiedmayer.
\newblock Quantum technologies with hybrid systems.
\newblock {\em PNAS}, 112(13):3866--3873, March 2015.

\bibitem{devoretSuperconductingCircuitsQuantum2013}
M.~H. Devoret and R.~J. Schoelkopf.
\newblock Superconducting {{Circuits}} for {{Quantum Information}}: {{An
  Outlook}}.
\newblock {\em Science}, 339(6124):1169--1174, March 2013.

\bibitem{afzeliusQuantumMemoryPhotons2015a}
Mikael Afzelius, Nicolas Gisin, and Hugues {de Riedmatten}.
\newblock Quantum memory for photons.
\newblock {\em Physics Today}, 68(12):42--47, November 2015.

\bibitem{pulizziSpintronics2012}
Fabio Pulizzi.
\newblock Spintronics.
\newblock {\em Nature Mater}, 11(5):367--367, May 2012.

\bibitem{slonczewskiCurrentdrivenExcitationMagnetic1996}
J.~C. Slonczewski.
\newblock Current-driven excitation of magnetic multilayers.
\newblock {\em Journal of Magnetism and Magnetic Materials}, 159(1):L1--L7,
  June 1996.

\bibitem{bhattiSpintronicsBasedRandom2017}
Sabpreet Bhatti, Rachid Sbiaa, Atsufumi Hirohata, Hideo Ohno, Shunsuke Fukami,
  and S.~N. Piramanayagam.
\newblock Spintronics based random access memory: A review.
\newblock {\em Materials Today}, 20(9):530--548, November 2017.

\bibitem{losbySpinMechanics2016}
Joseph~E. Losby and Mark~R. Freeman.
\newblock Spin {{Mechanics}}.
\newblock {\em arXiv:1601.00674 [cond-mat]}, January 2016.

\bibitem{wuNanocavityOptomechanicalTorque2017}
Marcelo Wu, Nathanael L.-Y. Wu, Tayyaba Firdous, Fatemeh Fani~Sani, Joseph~E.
  Losby, Mark~R. Freeman, and Paul~E. Barclay.
\newblock Nanocavity optomechanical torque magnetometry and radiofrequency
  susceptometry.
\newblock {\em Nature Nanotechnology}, 12(2):127--131, February 2017.

\bibitem{lachance-quirionHybridQuantumSystems2019a}
Dany {Lachance-Quirion}, Yutaka Tabuchi, Arnaud Gloppe, Koji Usami, and
  Yasunobu Nakamura.
\newblock Hybrid quantum systems based on magnonics.
\newblock {\em Appl. Phys. Express}, 12(7):070101, June 2019.

\bibitem{hueblHighCooperativityCoupled2013a}
Hans Huebl, Christoph~W. Zollitsch, Johannes Lotze, Fredrik Hocke, Moritz
  Greifenstein, Achim Marx, Rudolf Gross, and Sebastian T.~B. Goennenwein.
\newblock High {{Cooperativity}} in {{Coupled Microwave Resonator Ferrimagnetic
  Insulator Hybrids}}.
\newblock {\em Phys. Rev. Lett.}, 111(12):127003, September 2013.

\bibitem{tabuchiHybridizingFerromagneticMagnons2014a}
Yutaka Tabuchi, Seiichiro Ishino, Toyofumi Ishikawa, Rekishu Yamazaki, Koji
  Usami, and Yasunobu Nakamura.
\newblock Hybridizing {{Ferromagnetic Magnons}} and {{Microwave Photons}} in
  the {{Quantum Limit}}.
\newblock {\em Phys. Rev. Lett.}, 113(8):083603, August 2014.

\bibitem{zhangStronglyCoupledMagnons2014a}
Xufeng Zhang, Chang-Ling Zou, Liang Jiang, and Hong~X. Tang.
\newblock Strongly {{Coupled Magnons}} and {{Cavity Microwave Photons}}.
\newblock {\em Phys. Rev. Lett.}, 113(15):156401, October 2014.

\bibitem{haighDispersiveReadoutFerromagnetic2015a}
J.~A. Haigh, N.~J. Lambert, A.~C. Doherty, and A.~J. Ferguson.
\newblock Dispersive readout of ferromagnetic resonance for strongly coupled
  magnons and microwave photons.
\newblock {\em Phys. Rev. B}, 91(10):104410, March 2015.

\bibitem{soykalStrongFieldInteractions2010a}
{\"O}.~O. Soykal and M.~E. Flatt{\'e}.
\newblock Strong {{Field Interactions}} between a {{Nanomagnet}} and a
  {{Photonic Cavity}}.
\newblock {\em Phys. Rev. Lett.}, 104(7):077202, February 2010.

\bibitem{tabuchiCoherentCouplingFerromagnetic2015a}
Yutaka Tabuchi, Seiichiro Ishino, Atsushi Noguchi, Toyofumi Ishikawa, Rekishu
  Yamazaki, Koji Usami, and Yasunobu Nakamura.
\newblock Coherent coupling between a ferromagnetic magnon and a
  superconducting qubit.
\newblock {\em Science}, 349(6246):405--408, July 2015.

\bibitem{haighTripleResonantBrillouinLight2016}
J.~A. Haigh, A.~Nunnenkamp, A.~J. Ramsay, and A.~J. Ferguson.
\newblock Triple-{{Resonant Brillouin Light Scattering}} in
  {{Magneto}}-{{Optical Cavities}}.
\newblock {\em Physical Review Letters}, 117(13), September 2016.

\bibitem{zhangOptomagnonicWhisperingGallery2016}
Xufeng Zhang, Na~Zhu, Chang-Ling Zou, and Hong~X. Tang.
\newblock Optomagnonic {{Whispering Gallery Microresonators}}.
\newblock {\em Physical Review Letters}, 117(12), September 2016.

\bibitem{osadaCavityOptomagnonicsSpinOrbit2016}
A.~Osada, R.~Hisatomi, A.~Noguchi, Y.~Tabuchi, R.~Yamazaki, K.~Usami,
  M.~Sadgrove, R.~Yalla, M.~Nomura, and Y.~Nakamura.
\newblock Cavity {{Optomagnonics}} with {{Spin}}-{{Orbit Coupled Photons}}.
\newblock {\em Phys. Rev. Lett.}, 116(22):223601, June 2016.

\bibitem{violakusminskiyCoupledSpinlightDynamics2016}
Silvia Viola~Kusminskiy, Hong~X. Tang, and Florian Marquardt.
\newblock Coupled spin-light dynamics in cavity optomagnonics.
\newblock {\em Phys. Rev. A}, 94:033821, 2016.

\bibitem{liuOptomagnonicsMagneticSolids2016a}
Tianyu Liu, Xufeng Zhang, Hong~X. Tang, and Michael~E. Flatt{\'e}.
\newblock Optomagnonics in magnetic solids.
\newblock {\em Phys. Rev. B}, 94(6):060405, August 2016.

\bibitem{wangBistabilityCavityMagnon2018}
Yi-Pu Wang, Guo-Qiang Zhang, Dengke Zhang, Tie-Fu Li, C.-M. Hu, and J.~Q. You.
\newblock Bistability of {{Cavity Magnon Polaritons}}.
\newblock {\em Phys. Rev. Lett.}, 120(5):057202, January 2018.

\bibitem{maier-flaigTunableMagnonphotonCoupling2017a}
H.~{Maier-Flaig}, M.~Harder, S.~Klingler, Z.~Qiu, E.~Saitoh, M.~Weiler,
  S.~Gepr{\"a}gs, R.~Gross, S.~T.~B. Goennenwein, and H.~Huebl.
\newblock Tunable magnon-photon coupling in a compensating
  ferrimagnet\textemdash{}from weak to strong coupling.
\newblock {\em Appl. Phys. Lett.}, 110(13):132401, March 2017.

\bibitem{morrisStrongCouplingMagnons2017a}
R.~G.~E. Morris, A.~F. van Loo, S.~Kosen, and A.~D. Karenowska.
\newblock Strong coupling of magnons in a {{YIG}} sphere to photons in a planar
  superconducting resonator in the quantum limit.
\newblock {\em Sci Rep}, 7(1):1--6, September 2017.

\bibitem{boventerControlCouplingStrength2019}
Isabella Boventer, Christine D{\"o}rflinger, Tim Wolz, Rair Mac{\^e}do, Romain
  Lebrun, Mathias Kl{\"a}ui, and Martin Weides.
\newblock Control of the {{Coupling Strength}} and the {{Linewidth}} of a
  {{Cavity}}-{{Magnon Polariton}}.
\newblock {\em arXiv:1904.00393 [cond-mat]}, March 2019.

\bibitem{wangQuantumSimulationFermionBoson2019}
Yi-Pu Wang, Guo-Qiang Zhang, Da~Xu, Tie-Fu Li, Shi-Yao Zhu, J.~S. Tsai, and
  J.~Q. You.
\newblock Quantum {{Simulation}} of the {{Fermion}}-{{Boson Composite
  Quasi}}-{{Particles}} with a {{Driven Qubit}}-{{Magnon Hybrid Quantum
  System}}.
\newblock {\em arXiv:1903.12498 [cond-mat, physics:quant-ph]}, March 2019.

\bibitem{flowerExperimentalImplementationsCavitymagnon2019}
Graeme Flower, Maxim Goryachev, Jeremy Bourhill, and Michael~E. Tobar.
\newblock Experimental implementations of cavity-magnon systems: From ultra
  strong coupling to applications in precision measurement.
\newblock {\em New J. Phys.}, 21(9):095004, September 2019.

\bibitem{lachance-quirionResolvingQuantaCollective2017}
Dany {Lachance-Quirion}, Yutaka Tabuchi, Seiichiro Ishino, Atsushi Noguchi,
  Toyofumi Ishikawa, Rekishu Yamazaki, and Yasunobu Nakamura.
\newblock Resolving quanta of collective spin excitations in a millimeter-sized
  ferromagnet.
\newblock {\em Science Advances}, 3(7):e1603150, July 2017.

\bibitem{raoLevelAttractionLevel2019}
J.~W. Rao, C.~H. Yu, Y.~T. Zhao, Y.~S. Gui, X.~L. Fan, D.~S. Xue, and C.-M. Hu.
\newblock Level attraction and level repulsion of magnon coupled with a cavity
  anti-resonance.
\newblock {\em New J. Phys.}, 21(6):065001, June 2019.

\bibitem{boventerSteeringLevelRepulsion2019}
Isabella Boventer, Mathias Kl{\"a}ui, Rair Mac{\^e}do, and Martin Weides.
\newblock Steering between {{Level Repulsion}} and {{Attraction}}: {{Beyond
  Single}}-{{Tone Driven Cavity Magnon}}-{{Polaritons}}.
\newblock {\em arXiv:1908.05439 [cond-mat, physics:quant-ph]}, August 2019.

\bibitem{wangNonreciprocityUnidirectionalInvisibility2019}
Yi-Pu Wang, J.~W. Rao, Y.~Yang, Peng-Chao Xu, Y.~S. Gui, B.~M. Yao, J.~Q. You,
  and C.-M. Hu.
\newblock Nonreciprocity and {{Unidirectional Invisibility}} in {{Cavity
  Magnonics}}.
\newblock {\em Phys. Rev. Lett.}, 123(12):127202, September 2019.

\bibitem{lachance-quirionEntanglementbasedSingleshotDetection2019}
Dany {Lachance-Quirion}, Samuel~Piotr Wolski, Yutaka Tabuchi, Shingo Kono, Koji
  Usami, and Yasunobu Nakamura.
\newblock Entanglement-based single-shot detection of a single magnon with a
  superconducting qubit.
\newblock {\em arXiv:1910.09096 [cond-mat, physics:quant-ph]}, October 2019.

\bibitem{houStrongCouplingMicrowave2019}
Justin~T. Hou and Luqiao Liu.
\newblock Strong {{Coupling}} between {{Microwave Photons}} and {{Nanomagnet
  Magnons}}.
\newblock {\em Phys. Rev. Lett.}, 123(10):107702, September 2019.

\bibitem{liStrongCouplingMagnons2019}
Yi~Li, Tomas Polakovic, Yong-Lei Wang, Jing Xu, Sergi Lendinez, Zhizhi Zhang,
  Junjia Ding, Trupti Khaire, Hilal Saglam, Ralu Divan, John Pearson, Wai-Kwong
  Kwok, Zhili Xiao, Valentine Novosad, Axel Hoffmann, and Wei Zhang.
\newblock Strong {{Coupling}} between {{Magnons}} and {{Microwave Photons}} in
  {{On}}-{{Chip Ferromagnet}}-{{Superconductor Thin}}-{{Film Devices}}.
\newblock {\em Phys. Rev. Lett.}, 123(10):107701, September 2019.

\bibitem{cottamLightScatteringMagnetic1986}
Michael~G. Cottam and David~J. Lockwood.
\newblock {\em Light {{Scattering}} in {{Magnetic Solids}}}.
\newblock {Wiley-Interscience}, {New York}, 1 edition edition, August 1986.

\bibitem{pantazopoulosPhotomagnonicNanocavitiesStrong2017a}
P.~A. Pantazopoulos, N.~Stefanou, E.~Almpanis, and N.~Papanikolaou.
\newblock Photomagnonic nanocavities for strong light--spin-wave interaction.
\newblock {\em Phys. Rev. B}, 96(10):104425, September 2017.

\bibitem{grafCavityOptomagnonicsMagnetic2018a}
Jasmin Graf, Hannes Pfeifer, Florian Marquardt, and Silvia Viola~Kusminskiy.
\newblock Cavity optomagnonics with magnetic textures: {{Coupling}} a magnetic
  vortex to light.
\newblock {\em Phys. Rev. B}, 98(24):241406, December 2018.

\bibitem{sharmaOptimalModeMatching2019}
Sanchar Sharma, Babak~Zare Rameshti, Yaroslav~M. Blanter, and Gerrit E.~W.
  Bauer.
\newblock Optimal mode matching in cavity optomagnonics.
\newblock {\em Phys. Rev. B}, 99(21):214423, June 2019.

\bibitem{pantazopoulosTailoringCouplingLight2018}
Petros-Andreas Pantazopoulos, Nikolaos Papanikolaou, and Nikolaos Stefanou.
\newblock Tailoring coupling between light and spin waves with dual
  photonic\textendash{}magnonic resonant layered structures.
\newblock {\em J. Opt.}, 21(1):015603, December 2018.

\bibitem{almpanisDielectricMagneticMicroparticles2018}
Evangelos Almpanis.
\newblock Dielectric magnetic microparticles as photomagnonic cavities:
  {{Enhancing}} the modulation of near-infrared light by spin waves.
\newblock {\em Phys. Rev. B}, 97(18):184406, May 2018.

\bibitem{osadaOrbitalAngularMomentum2018}
A.~Osada, A.~Gloppe, Y.~Nakamura, and K.~Usami.
\newblock Orbital angular momentum conservation in {{Brillouin}} light
  scattering within a ferromagnetic sphere.
\newblock {\em New J. Phys.}, 20(10):103018, October 2018.

\bibitem{osadaBrillouinLightScattering2018}
A.~Osada, A.~Gloppe, R.~Hisatomi, A.~Noguchi, R.~Yamazaki, M.~Nomura,
  Y.~Nakamura, and K.~Usami.
\newblock Brillouin {{Light Scattering}} by {{Magnetic Quasivortices}} in
  {{Cavity Optomagnonics}}.
\newblock {\em Phys. Rev. Lett.}, 120(13):133602, March 2018.

\bibitem{haighSelectionRulesCavityenhanced2018a}
J.~A. Haigh, N.~J. Lambert, S.~Sharma, Y.~M. Blanter, G.~E.~W. Bauer, and A.~J.
  Ramsay.
\newblock Selection rules for cavity-enhanced {{Brillouin}} light scattering
  from magnetostatic modes.
\newblock {\em Phys. Rev. B}, 97(21):214423, June 2018.

\bibitem{pantazopoulosHighefficiencyTripleresonantInelastic2019}
Petros~Andreas Pantazopoulos, Kosmas~L. Tsakmakidis, Evangelos Almpanis,
  Grigorios~P. Zouros, and Nikolaos Stefanou.
\newblock High-efficiency triple-resonant inelastic light scattering in planar
  optomagnonic cavities.
\newblock {\em New J. Phys.}, 21(9):095001, September 2019.

\bibitem{hisatomiHelicityChangingBrillouinLight2019a}
R.~Hisatomi, A.~Noguchi, R.~Yamazaki, Y.~Nakata, A.~Gloppe, Y.~Nakamura, and
  K.~Usami.
\newblock Helicity-changing brillouin light scattering by magnons in a
  ferromagnetic crystal.
\newblock {\em Phys. Rev. Lett.}, 123:207401, Nov 2019.

\bibitem{purcelle.mProceedingsAmericanPhysical1946}
{Purcell, E. M}.
\newblock Proceedings of the {{American Physical Society}}.
\newblock {\em Physical Review}, 69(11-12):674--674, June 1946.

\bibitem{harocheCavityQuantumElectrodynamics1989}
Serge Haroche and Daniel Kleppner.
\newblock Cavity {{Quantum Electrodynamics}}.
\newblock {\em Physics Today}, 42(1):24--30, January 1989.

\bibitem{waltherCavityQuantumElectrodynamics2006}
Herbert Walther, Benjamin T.~H. Varcoe, Berthold-Georg Englert, and Thomas
  Becker.
\newblock Cavity quantum electrodynamics.
\newblock {\em Rep. Prog. Phys.}, 69(5):1325--1382, April 2006.

\bibitem{schoelkopfWiringQuantumSystems2008}
R.~J. Schoelkopf and S.~M. Girvin.
\newblock Wiring up quantum systems.
\newblock {\em Nature}, 451(7179):664--669, February 2008.

\bibitem{stancilSpinWavesTheory2009}
Daniel~D. Stancil and Anil Prabhakar.
\newblock {\em Spin {{Waves}}: {{Theory}} and {{Applications}}}.
\newblock {Springer US}, 2009.

\bibitem{l.d.landauElectrodynamicsContinuousMedia}
{L. D. Landau}, {L. P. Pitaevskii}, and {E. M. Lifshitz}.
\newblock {\em Electrodynamics of {{Continuous Media}}: {{Volume}} 8
  ({{Course}} of {{Theoretical Physics S}})}.
\newblock {Butterworth-Heinemann}, 2nd edition.

\bibitem{kusminiskiyQuantumMagnetismSpin2019}
Silvia~Viola Kusminiskiy.
\newblock {\em Quantum {{Magnetism}}, {{Spin Waves}}, and {{Optical
  Cavities}}}.
\newblock {{SpringerBriefs}} in {{Physics}}. {Springer International
  Publishing}, 2019.

\bibitem{garstCollectiveSpinExcitations2017a}
Markus Garst, Johannes Waizner, and Dirk Grundler.
\newblock Collective spin excitations of helices and magnetic skyrmions: Review
  and perspectives of magnonics in non-centrosymmetric magnets.
\newblock {\em Journal of Physics D: Applied Physics}, 50(29):293002, July
  2017.

\bibitem{dennisDefiningLengthScales2002}
C.~L. Dennis, R.~P. Borges, L.~D. Buda, U.~Ebels, J.~F. Gregg, M.~Hehn,
  E.~Jouguelet, K.~Ounadjela, I.~Petej, I.~L. Prejbeanu, and M.~J. Thornton.
\newblock The defining length scales of mesomagnetism: A review.
\newblock {\em J. Phys.: Condens. Matter}, 14(49):R1175, 2002.

\bibitem{kittelPhysicalTheoryFerromagnetic1949}
Charles Kittel.
\newblock Physical {{Theory}} of {{Ferromagnetic Domains}}.
\newblock {\em Rev. Mod. Phys.}, 21(4):541--583, October 1949.

\bibitem{heebnerOpticalMicroresonatorsTheory2008}
John Heebner, Rohit Grover, and Tarek Ibrahim.
\newblock {\em Optical {{Microresonators}}: {{Theory}}, {{Fabrication}}, and
  {{Applications}}}.
\newblock Springer {{Series}} in {{Optical Sciences}}. {Springer-Verlag}, {New
  York}, 2008.

\bibitem{aspelmeyerCavityOptomechanicsNano2014}
Markus Aspelmeyer, Tobias~J. Kippenberg, and Florian Marquardt, editors.
\newblock {\em Cavity {{Optomechanics}}: {{Nano}}- and {{Micromechanical
  Resonators Interacting}} with {{Light}}}.
\newblock Quantum {{Science}} and {{Technology}}. {Springer-Verlag}, {Berlin
  Heidelberg}, 2014.

\bibitem{maccabePhononicBandgapNanoacoustic2019}
Gregory~S. MacCabe, Hengjiang Ren, Jie Luo, Justin~D. Cohen, Hengyun Zhou, Alp
  Sipahigil, Mohammad Mirhosseini, and Oskar Painter.
\newblock Phononic bandgap nano-acoustic cavity with ultralong phonon lifetime.
\newblock {\em arXiv:1901.04129 [cond-mat, physics:quant-ph]}, January 2019.

\bibitem{sharmaLightScatteringMagnons2017}
Sanchar Sharma, Yaroslav~M. Blanter, and Gerrit E.~W. Bauer.
\newblock Light scattering by magnons in whispering gallery mode cavities.
\newblock {\em Phys. Rev. B}, 96(9):094412, September 2017.

\bibitem{osbornDemagnetizingFactorsGeneral1945}
J.~A. Osborn.
\newblock Demagnetizing {{Factors}} of the {{General Ellipsoid}}.
\newblock {\em Phys. Rev.}, 67(11-12):351--357, June 1945.

\bibitem{wallsQuantumOptics2008}
D.~F. Walls and Gerard~J. Milburn.
\newblock {\em Quantum {{Optics}}}.
\newblock {Springer-Verlag}, {Berlin Heidelberg}, 2 edition, 2008.

\bibitem{meystreElementsQuantumOptics2007}
Pierre Meystre and Murray Sargent.
\newblock {\em Elements of {{Quantum Optics}}}.
\newblock {Springer-Verlag}, {Berlin Heidelberg}, 4th edition, 2007.

\bibitem{clogstonFerromagneticResonanceLine1956}
A.~M. Clogston, H.~Suhl, L.~R. Walker, and P.~W. Anderson.
\newblock Ferromagnetic resonance line width in insulating materials.
\newblock {\em Journal of Physics and Chemistry of Solids}, 1(3):129--136,
  November 1956.

\bibitem{kohlerCavityAssistedMeasurementCoherent2017}
Jonathan Kohler, Nicolas Spethmann, Sydney Schreppler, and Dan~M.
  {Stamper-Kurn}.
\newblock Cavity-{{Assisted Measurement}} and {{Coherent Control}} of
  {{Collective Atomic Spin Oscillators}}.
\newblock {\em Phys. Rev. Lett.}, 118(6):063604, February 2017.

\bibitem{marquardtDynamicalMultistabilityInduced2006}
Florian Marquardt, J.~G.~E. Harris, and S.~M. Girvin.
\newblock Dynamical {{Multistability Induced}} by {{Radiation Pressure}} in
  {{High}}-{{Finesse Micromechanical Optical Cavities}}.
\newblock {\em Phys. Rev. Lett.}, 96(10):103901, March 2006.

\bibitem{guslienkoMagneticVortexState2008}
K.~Yu. Guslienko.
\newblock Magnetic {{Vortex State Stability}}, {{Reversal}} and {{Dynamics}} in
  {{Restricted Geometries}}.
\newblock {\em Journal of Nanoscience and Nanotechnology}, 8(6):2745--2760,
  June 2008.

\bibitem{pigeauOptimalControlVortexcore2011}
Benjamin Pigeau, Gr{\'e}goire {de Loubens}, Olivier Klein, Andreas Riegler,
  Florian Lochner, Georg Schmidt, and Laurens~W. Molenkamp.
\newblock Optimal control of vortex-core polarity by resonant microwave pulses.
\newblock {\em Nat Phys}, 7(1):26--31, January 2011.

\bibitem{thieleSteadyStateMotionMagnetic1973}
A.~A. Thiele.
\newblock Steady-{{State Motion}} of {{Magnetic Domains}}.
\newblock {\em Phys. Rev. Lett.}, 30(6):230--233, February 1973.

\bibitem{vansteenkisteDesignVerificationMuMax32014}
Arne Vansteenkiste, Jonathan Leliaert, Mykola Dvornik, Mathias Helsen, Felipe
  {Garcia-Sanchez}, and Bartel Van~Waeyenberge.
\newblock The design and verification of {{MuMax3}}.
\newblock {\em AIP Advances}, 4(10):107133, October 2014.

\bibitem{bittencourtMagnonHeraldingCavity2019}
Victor A. S.~V. Bittencourt, Verena Feulner, and Silvia~Viola Kusminskiy.
\newblock Magnon heralding in cavity optomagnonics.
\newblock {\em Phys. Rev. A}, 100(1):013810, July 2019.

\bibitem{sharmaOpticalCoolingMagnons2018}
Sanchar Sharma, Yaroslav~M. Blanter, and Gerrit E.~W. Bauer.
\newblock Optical cooling of magnons.
\newblock {\em Phys. Rev. Lett.}, 121:087205, Aug 2018.

\end{thebibliography}

\end{document}